\documentclass[twocolumn]{aastex631}

\usepackage{times}
\usepackage{amsmath,amssymb,amstext}
\hypersetup{linkcolor=red,citecolor=blue,filecolor=cyan,urlcolor=blue}
\usepackage{tabularx}
\usepackage{longtable}
\usepackage{float}
\usepackage{natbib}
\usepackage{graphicx,color}
\usepackage{txfonts}

\usepackage{multirow}
\usepackage{array}
\usepackage{rotating}
\usepackage{sidecap}
\usepackage{hyperref}
\usepackage{epstopdf}
\usepackage{footnote}
\usepackage{tabularx}
\usepackage{booktabs}
\usepackage{longtable}
\usepackage{tabu}
\usepackage{longtable}

\newcommand{\kms}{km\,s$^{-1}$}

\begin{document}

\title{{\large {Direct Imaging of a Prolonged Plasma/Current Sheet and Quasiperiodic Magnetic Reconnection on the Sun}}}
\author{Pankaj Kumar\altaffiliation{1,2}}

\affiliation{Department of Physics, American University, Washington, DC 20016, USA}
\affiliation{Heliophysics Science Division, NASA Goddard Space Flight Center, Greenbelt, MD, 20771, USA}

\author{Judith T.\ Karpen}
\affiliation{Heliophysics Science Division, NASA Goddard Space Flight Center, Greenbelt, MD, 20771, USA}

\author{Vasyl Yurchyshyn}
\affiliation{Big Bear Solar Observatory, New Jersey Institute of Technology, Big Bear City, CA, 92314, USA}

\author{C.\ Richard DeVore}
\affiliation{Heliophysics Science Division, NASA Goddard Space Flight Center, Greenbelt, MD, 20771, USA}

\author{Spiro K. Antiochos}
\affiliation{Department of Climate and Space Sciences and Engineering, University of Michigan, Ann Arbor, MI 48109, USA}

\email{pankaj.kumar@nasa.gov}

\begin{abstract}
Magnetic reconnection is widely believed to be the fundamental process in the solar atmosphere that underlies magnetic energy release and particle acceleration. This process is responsible for the onset of solar flares, coronal mass ejections, and other explosive events (e.g., jets).
Here, we report direct imaging of a prolonged plasma/current sheet along with quasiperiodic magnetic reconnection in the solar corona using ultra-high-resolution observations from the 1.6-meter Goode Solar Telescope (GST) at BBSO and Solar Dynamics Observatory/Atmospheric Imaging Assembly (SDO/AIA). The current sheet appeared near a null point in the fan-spine topology and persisted over an extended period ($\approx$20 hours). The length and apparent width of the current sheet were about 6$\arcsec$ and 2$\arcsec$ respectively, and the plasma temperature was $\approx$10-20 MK. We observed quasiperiodic plasma inflows and outflows (bidirectional jets with plasmoids) at the reconnection site/current sheet. Furthermore, quasiperiodic reconnection at the long-lasting current sheet produced recurrent eruptions (small flares and jets) and contributed significantly to the recurrent impulsive heating of the active region. Direct imaging of a plasma/current sheet and recurrent null-point reconnection for such an extended period has not been reported previously. These unprecedented observations provide compelling evidence that supports the universal model for solar eruptions (i.e., the breakout model) and have implications for impulsive heating of active regions by recurrent reconnection near null points. The prolonged and sustained reconnection for about 20 hours at the breakout current sheet provides new insights into the dynamics and energy release processes in the solar corona.
\end{abstract}
\section{INTRODUCTION}\label{intro}
Magnetic reconnection is a fundamental process in the Sun's atmosphere that drives the release of magnetic energy and the acceleration of particles. This dynamic phenomenon plays a crucial role in various explosive events, including solar flares, coronal mass ejections, and jets \citep{shibata2011,chen2011,raouafi2016}. Gaining insights into the mechanisms and features of magnetic reconnection is crucial for a comprehensive understanding of solar energetic events, associated particle acceleration, heating, and their influence on space weather. 

Coronal bright points (i.e., small active regions) are ubiquitous on the Sun and can be found in coronal holes, the quiet Sun, and active regions \citep{golub1977,madjarska2019,raouafi2023}. They are the source regions for small-scale eruptions and jets \citep{shibata2007, sterling2015}, and may play a crucial role in supplying mass and energy flux to the solar corona and solar wind. The majority of these bright points exhibit fan-spine topologies with a null point \citep{galsgaard2017,kumar2019a}.
The magnetic breakout model for solar eruptions explains solar activities ranging from coronal mass ejections (CMEs) to coronal jets, through magnetic reconnection at the null point  \citep{antiochos1998,antiochos1999,karpen2012,guidoni2016}. According to the breakout model of coronal jets \citep{wyper2017, wyper2018}, the eruption of filament channels, powered by free energy in the stressed field, first leads to slowly rising channel flux and the formation of a breakout current sheet (BCS) near the null point. Gradual breakout reconnection at the BCS removes restraining flux, enabling more expansion and the formation of a flare current sheet (FCS) under the rising filament-channel flux. Slow flare reconnection at the vertical FCS builds a rising flux rope, which drives explosive breakout reconnection when it reaches the BCS. This interaction destroys the flux rope and generates a helical jet along the outer spine. The plasma sheets surrounding both current sheets (BCS and FCS), along with multiple plasmoids, have been detected in coronal-hole and active-region jets originating from fan-spine topologies \citep{kumar2018,kumar2019b}, supporting the breakout paradigm for such events. Moreover, several case studies have shown evidence for plasma sheets during flares, eruptions, and jets – e.g., the large-scale vertical current sheet (CS) below the erupting flux rope in the 2017 September 10 flare \citep{cheng2018,chen2020}; CSs  undergoing reconnection \citep{li2016,xue2018,li2021,yan2022,cheng2023}; and BCS and FCS in failed flux rope eruptions \citep{kumar2023a,karpen2024}. \citet{ghosh2020} reported evidence of reconnection inflows/outflows at an X-point between large-scale loops connecting two active regions. In \S3 we demonstrate that the observations presented in \S2 are consistent with both gradual and explosive phases of the breakout model. 

Quasiperiodic reconnection jets have been observed emanating from fan-spine configurations in coronal holes \citep{kumar2019a,kumar2022,kumar2023}. However, determining the exact trigger of null-point reconnection has proven challenging due to the small size of bright points and the limited resolution of SDO/AIA. This naturally raises key questions about the greater significance of recurrent reconnection jets. What mechanisms lead to the accumulation of sufficient magnetic energy to power the quasiperiodic jets? How much can these jets contribute to heating the solar corona? Are the frequency and energy flux of reconnection jets sufficient to heat the corona? What is the typical temperature of the heated plasma during these reconnection jets? These are the crucial questions that we aim to address here.

A few different reconnection-based processes have been proposed for heating the active regions: nanoflares (\citealt{parker1988,klimchuk2015}; energy=10$^{24}$ ergs), type II spicules \citep{dePontieu2007,depontieu2011}; and X-ray microflares (\citealt{lin1984}; energy=10$^{26}$ to 10$^{28}$) have been speculated as the predominant source of active-region coronal heating \citep{sakurai2017}. High-resolution observations offer an excellent opportunity to determine, if any, which of these mechanisms is the primary contributor to coronal heating.

In most studies of jets in fan-spine topologies thus far, breakout CSs appear for short durations and disappear shortly after the jet. In this paper, we present direct imaging of a breakout plasma sheet and evidence for sustained quasiperiodic magnetic reconnection near a magnetic null point for about 20 hours. The recurrent reconnection continues until the fan-spine topology decays. 
Utilizing data from the Atmospheric Imaging Assembly (AIA) aboard the Solar Dynamics Observatory (SDO) and the 1.6-meter Goode Solar Telescope (GST) at the Big Bear Solar Observatory (BBSO), we detected quasiperiodic magnetic reconnection within the current sheet, producing eruptions and jets. Our observations reveal plasma inflows associated with bidirectional outflows and plasmoids above the reconnection site. These findings indicate the ongoing recurrent magnetic reconnection and dynamic energy release. Furthermore, the quasiperiodic nature of the reconnection events within the long-lasting current sheet near the null point is identified as a key factor contributing to repeated flares, eruptions, and jets, contributing significant mass and energy flux to the active region. 
 The recurrent impulsive heating of the active region associated with quasiperiodic reconnection events challenges existing models and introduces new perspectives on the dynamical processes governing the solar corona. These unprecedented observations not only confirm the validity of the breakout model for solar eruptions but also explain the observed quasiperiodic pulsations as evidence for recurrent reconnection at the null-point current sheet.

\begin{figure*}
\centering{
\includegraphics[width=9.6cm]{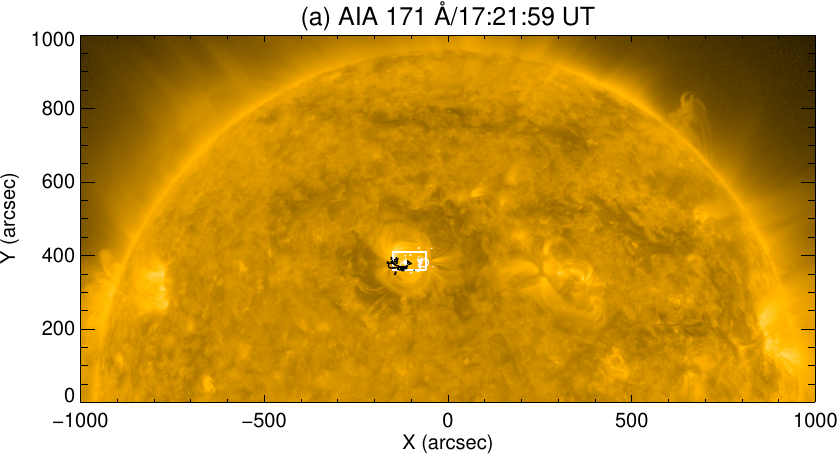}
\includegraphics[width=8.3cm]{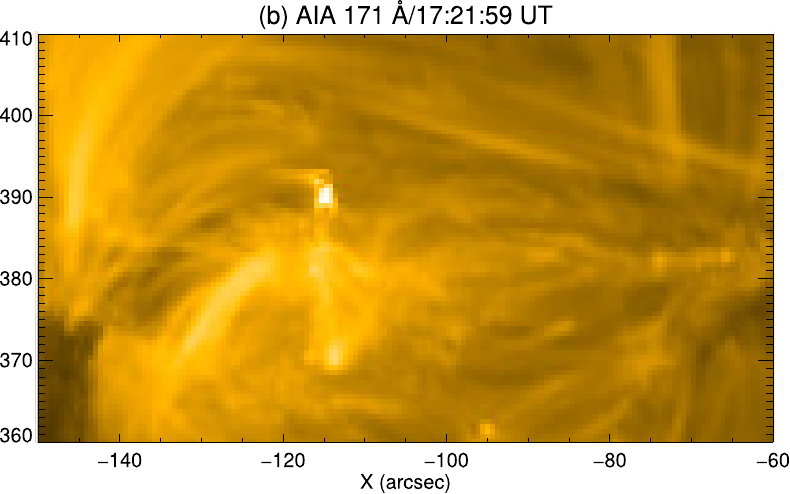}

\includegraphics[width=9.1cm]{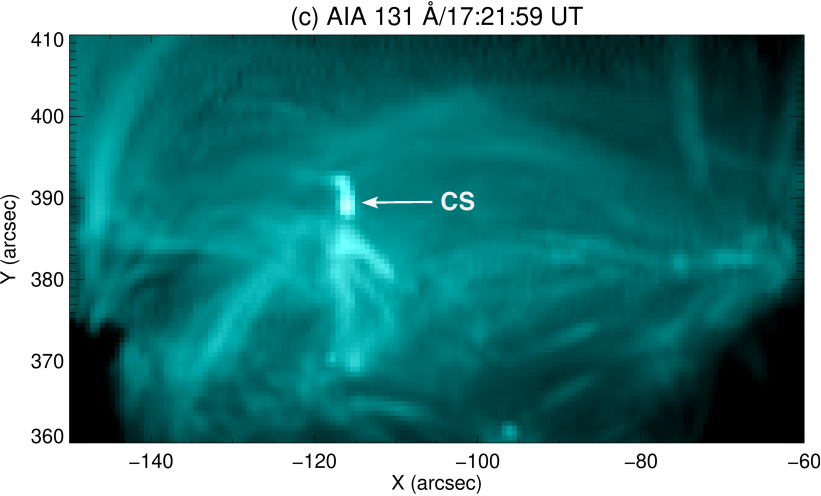}
\includegraphics[width=8.7cm]{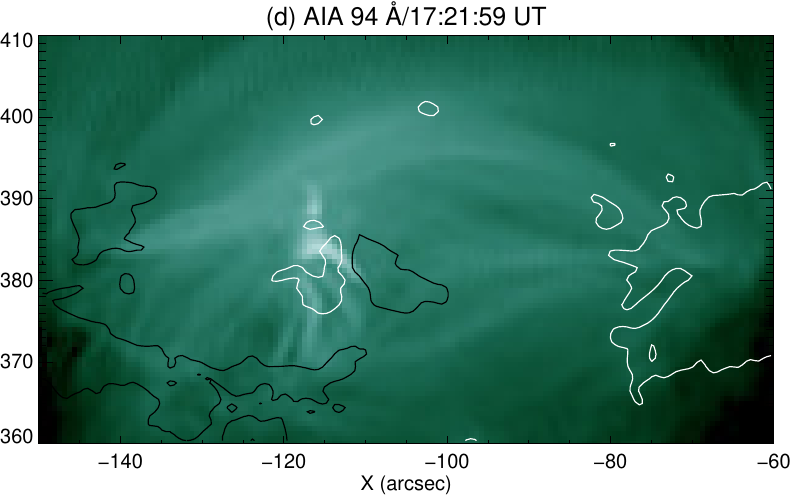}
\includegraphics[width=18cm]{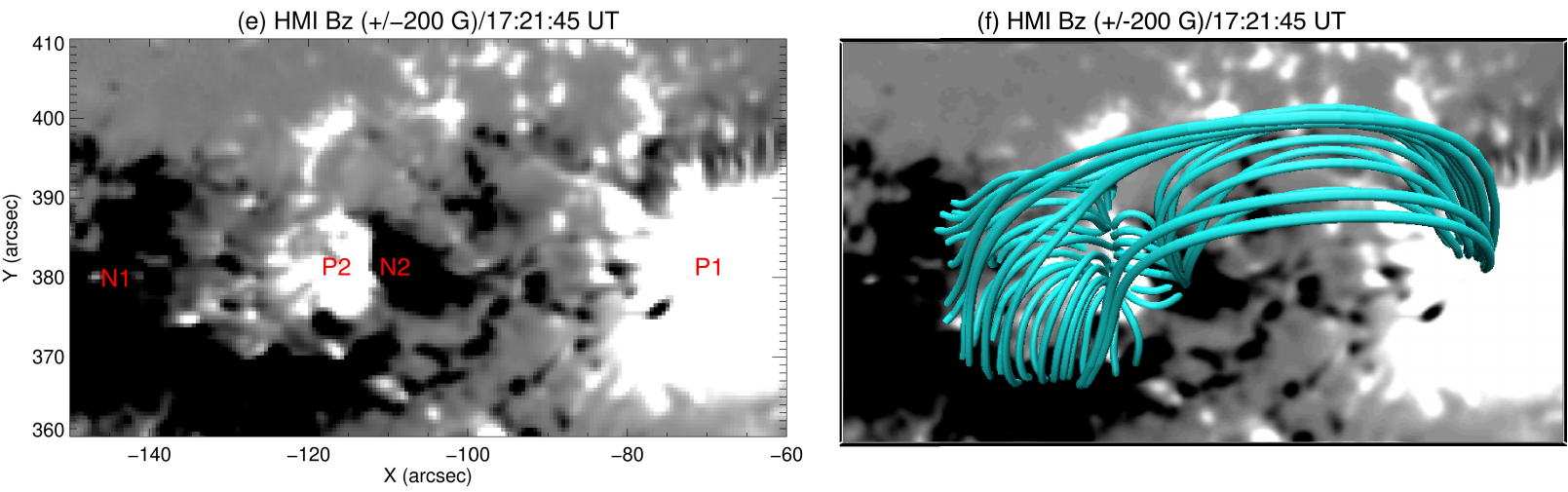}
}
\caption{{\bf Magnetic configuration of the flare/eruption site.} (a,b) AIA 171 {\AA} images of the eruption site on August 22, 2022. The zoomed view of the AR (white box in (a)) is shown in (b). (c,d) AIA 131 and 94 {\AA} images of the eruption site. An arrow indicates the current sheet (marked by ``CS") appeared near the null point. AIA 171 and 94 {\AA} images are overlaid by cotemporal HMI magnetogram contours ($\pm$500 G) of positive (white) and negative (black) polarities. (e,f) HMI line of sight magnetogram (Bz=$\pm$200 G) and potential field extrapolation of the source region. N1, P1, N2, and P2 indicate opposite polarities (negative/positive) in the quadrupolar magnetic configuration. The central fan-spine topology is clearest in panels (c) and (f).} 
\label{fig1}
\end{figure*}


\begin{figure*}
\centering{
\includegraphics[width=8.9cm]{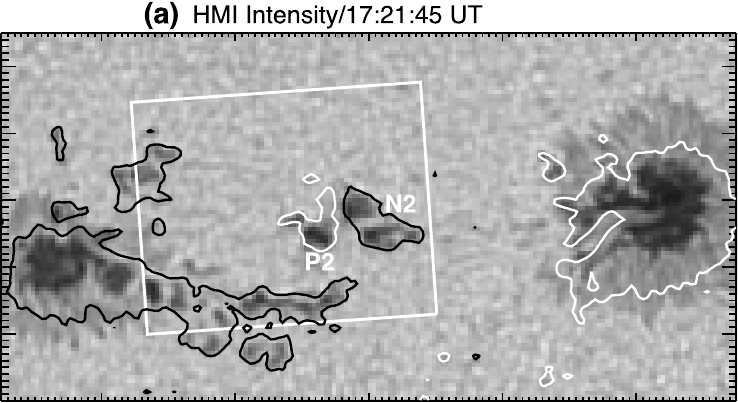}
\includegraphics[width=8.9cm]{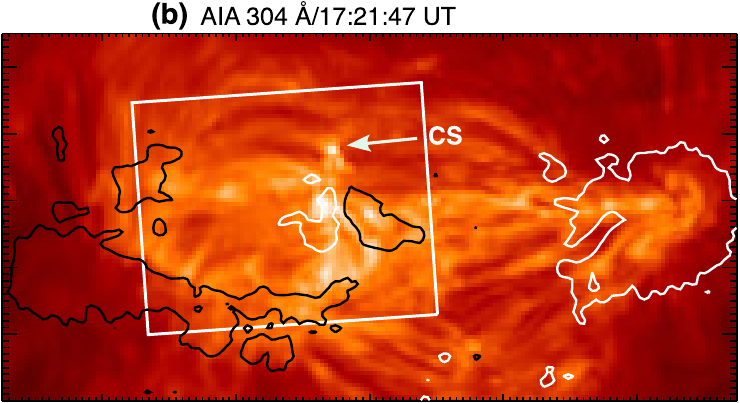}
\includegraphics[width=8.9cm]{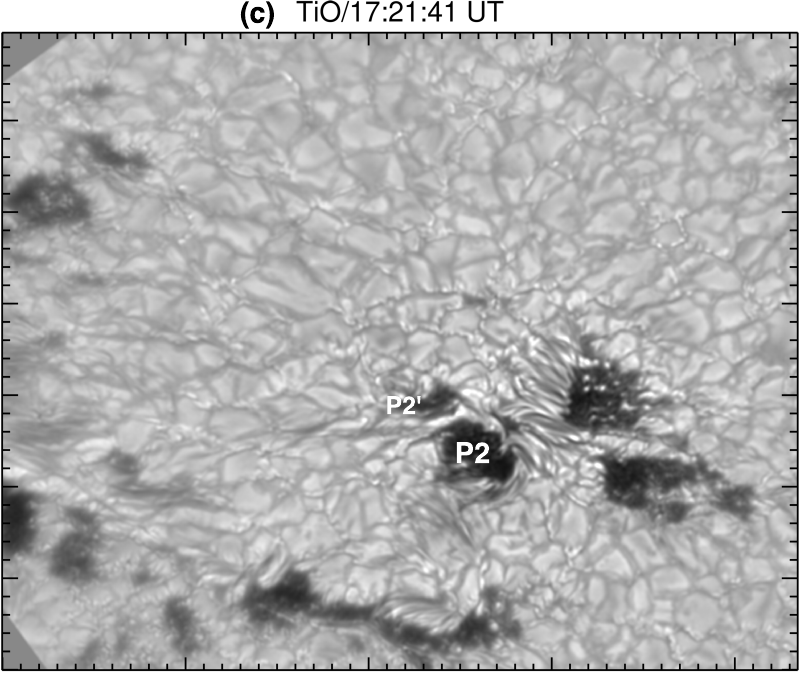}
\includegraphics[width=8.9cm]{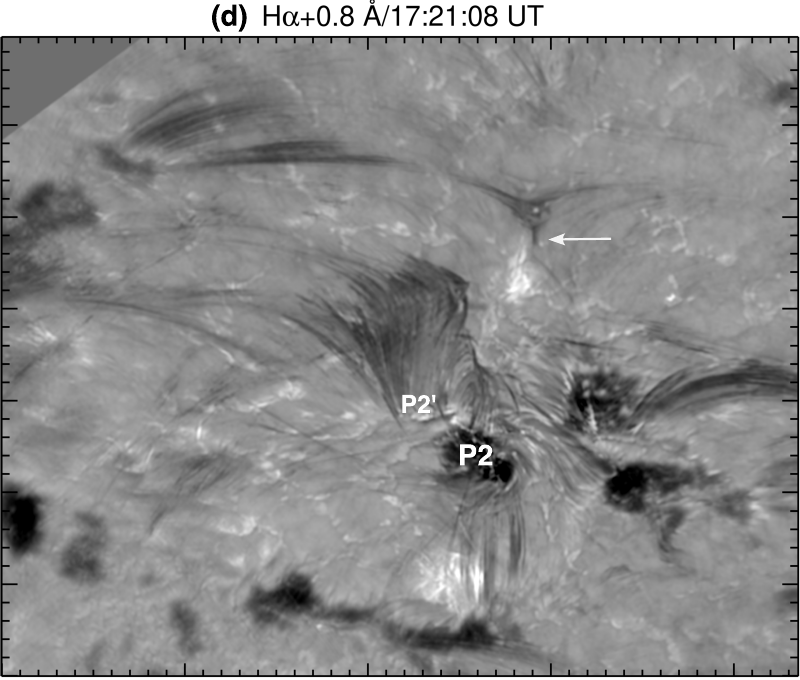}
\includegraphics[width=8.9cm]{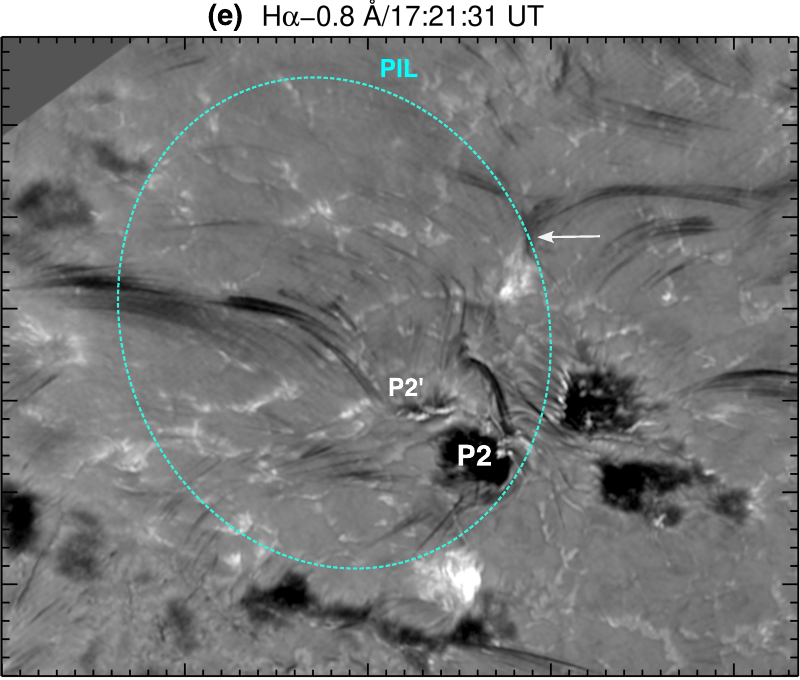}
\includegraphics[width=8.9cm]{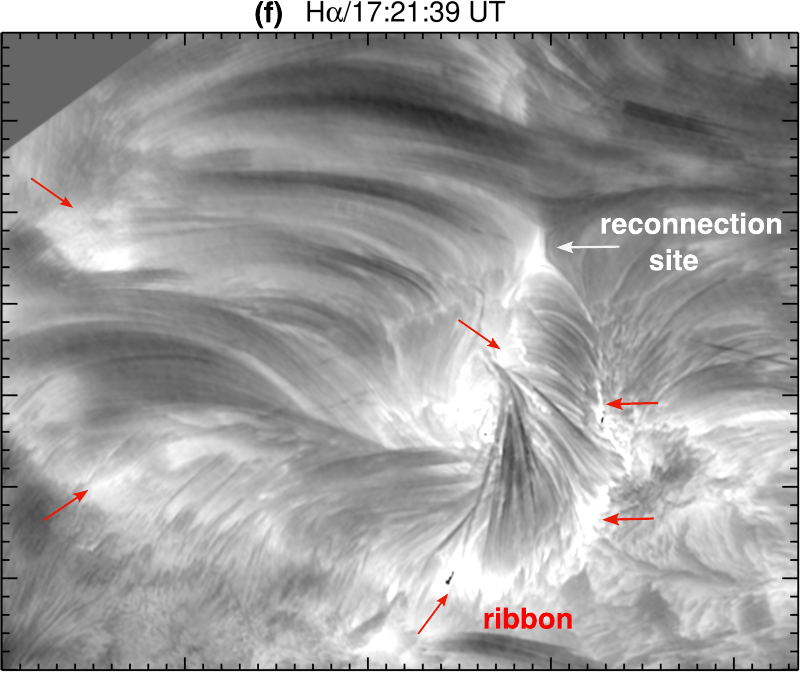}

}
\caption{{\bf GST images of the plasma sheet and surroundings.} (a,b) SDO/HMI continuum and AIA 304 {\AA} images overlaid by HMI magnetogram contours ($\pm$500 G, white=positive, black=negative). The field of view of co-aligned GST images is marked by a rectangle. An arrow indicates the bright plasma around the current sheet (marked by ``CS"). (c-f) Photospheric TiO image showing the zoomed view of the base (sunspots) of the fan-spine topology. H$\alpha$ ($\pm 0.8$ {\AA}, line center) images during magnetic reconnection show the structure rooted in moving magnetic features and associated inflows/brightenings at the footpoint of fan loops. P2$^\prime$ is the moving magnetic feature/spot that merges with P2. Both have the same polarity (positive). The reconnection site is indicated by white arrows, while a quasicircular/flare ribbons are depicted by red arrows. The approximate position of the quasicircular PIL is marked by a cyan dashed line. Each division on both axes corresponds to a distance of 1$\arcsec$.} 
\label{fig2}
\end{figure*}

\begin{figure*}
\centering{
\includegraphics[width=7.5cm]{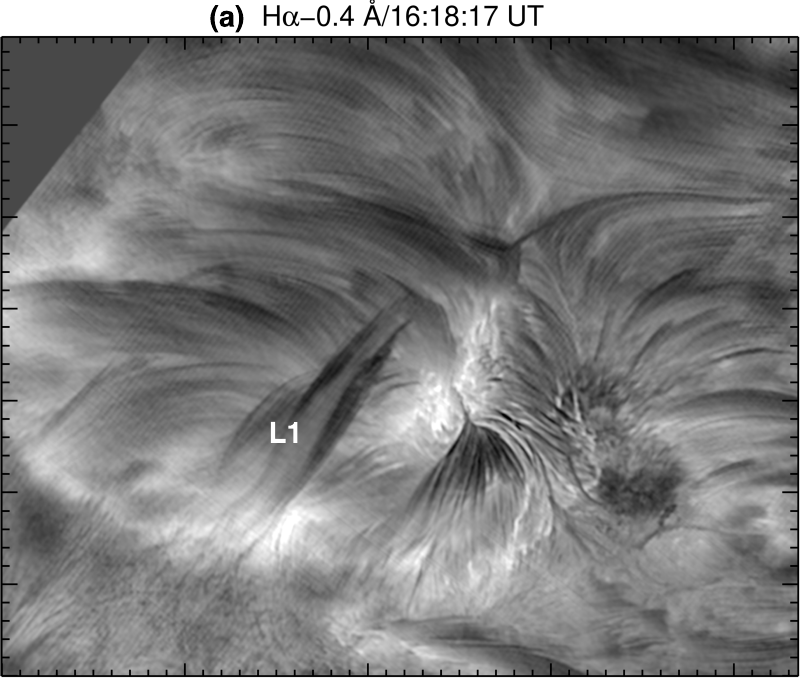}
\includegraphics[width=7.5cm]{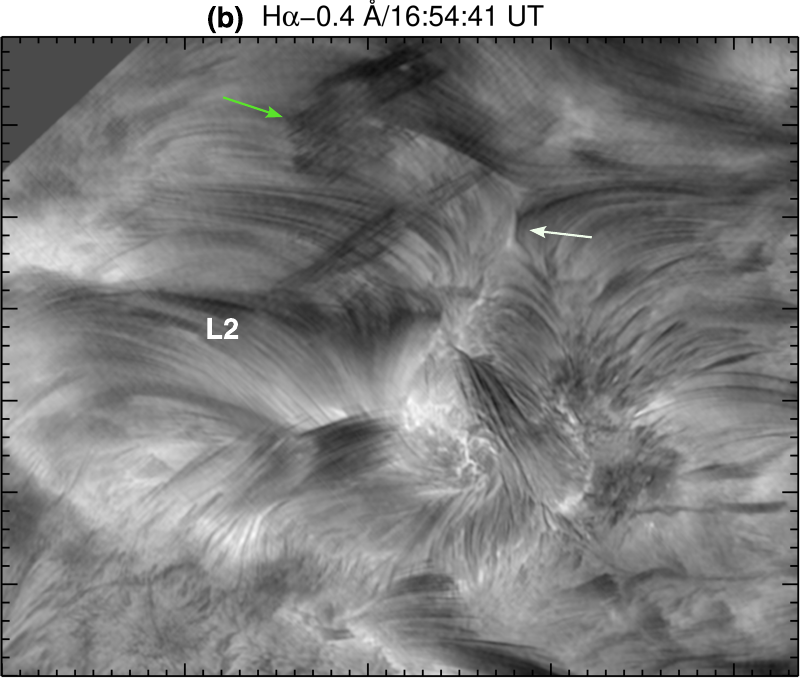}

\includegraphics[width=7.5cm]{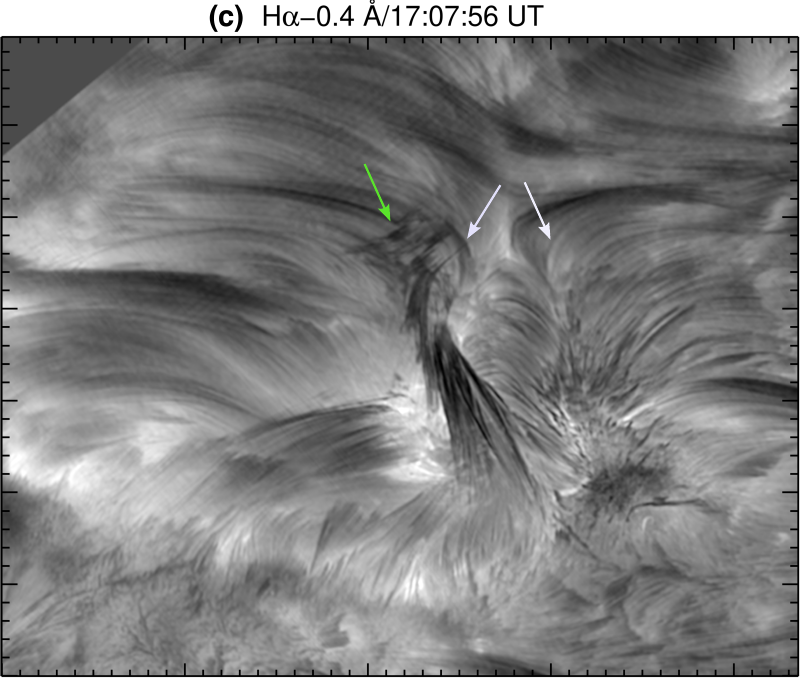}
\includegraphics[width=7.5cm]{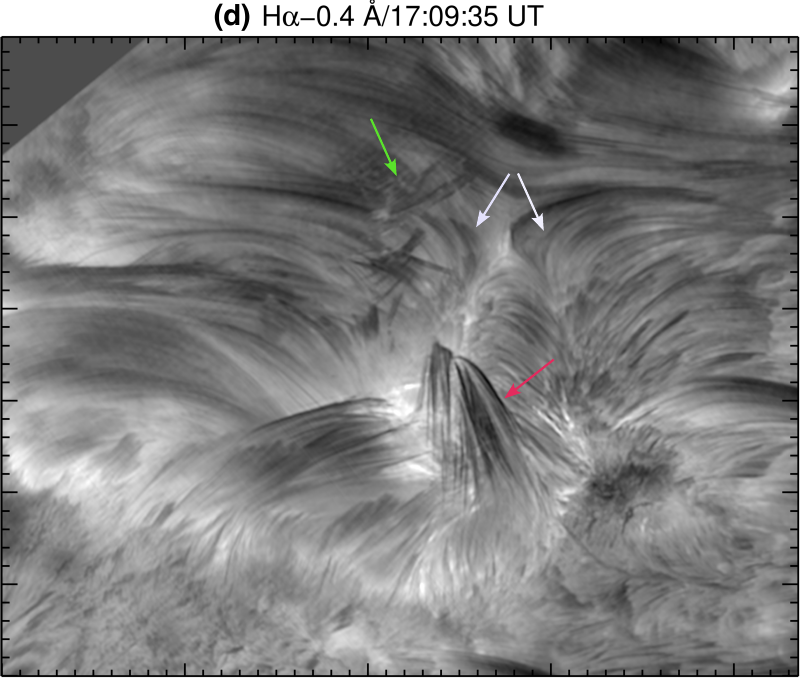}

\includegraphics[width=7.5cm]{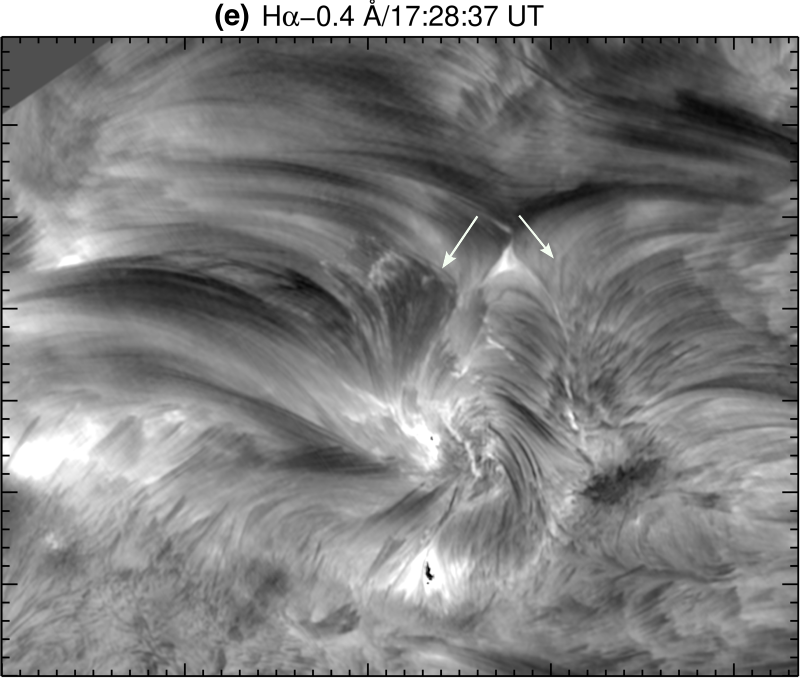}
\includegraphics[width=7.5cm]{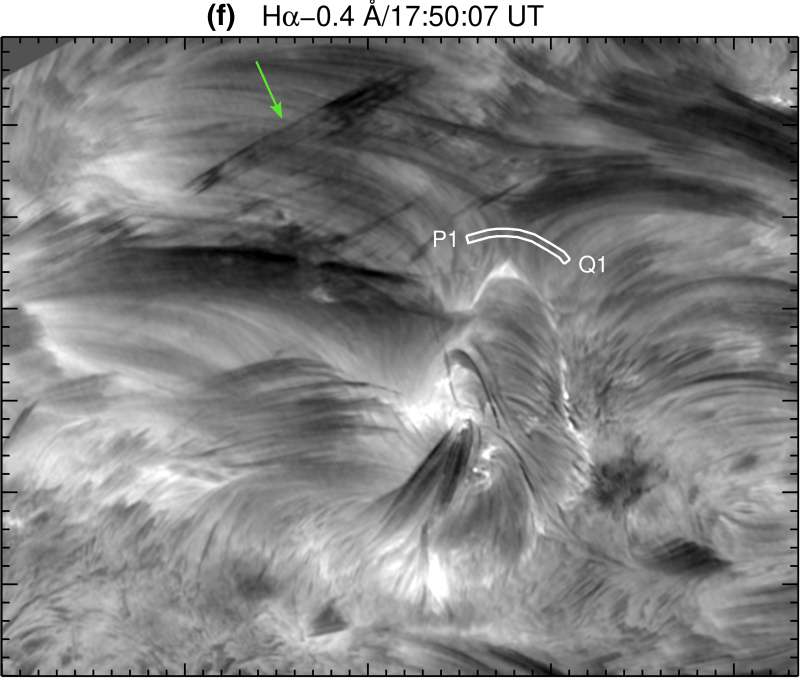}
}
\caption{{\bf Inflows and outflows associated with magnetic reconnection.} H$\alpha$-0.4 {\AA} images showing multiple structures (L1, L2) rising from the left side of the dome. These structures progress toward the current sheet, undergo reconnection, and subsequently produce outflows/jets. P1Q1 (curved slice) is used to construct the time-distance intensity plot in Figure \ref{fig5}(a). Green arrows = outflows, white arrows = inflows toward the preexisting CS, and red arrows=reconnection- associated flux transfer (closing).  Each division on both axes corresponds to a distance of 1$\arcsec$.} 
\label{fig3}
\end{figure*}

\section{OBSERVATIONS and RESULTS}\label{obs}
We analyzed GST (Goode Solar Telescope: \citealt{cao2010,goode2010}) images captured between 16:15 and 18:00 UT on August 22, 2022. GST offers high-resolution images of the photosphere and chromosphere. Specifically, we utilized the Visible Imaging Spectrometer (VIS) data (H$\alpha$, H$\alpha\pm$0.08 {\AA}, H$\alpha\pm$0.04 {\AA}) with a field of view (FOV) of 75$\arcsec$$\times$64$\arcsec$ and a pixel scale of 0.029$\arcsec$. The per pixel resolution for photospheric TiO images (7057 \AA) is 0.034$\arcsec$.

We analyzed SDO/AIA \citep{lemen2012} full-disk images of the Sun (FOV $\approx$ 1.3~$R_\sun$) with a spatial resolution of 1.5$\arcsec$ (0.6$\arcsec$~pixel$^{-1}$) and a cadence of 12~s, in the following channels: 304 ~{\AA} (\ion{He}{2}, $T\approx 0.05$~MK), 171~{\AA} (\ion{Fe}{9}, $T\approx 0.7$~MK), 193~{\AA} (\ion{Fe}{12}, \ion{Fe}{24}, $T\approx  1.2$~MK and $\approx 20$~MK), 211~{\AA} (\ion{Fe}{14}, $T\approx 2$~MK), 94~{\AA} (\ion{Fe}{10}, \ion{Fe}{18}, $T\approx 1, 6.3$~MK), and 131~\AA\ (\ion{Fe}{8}, \ion{Fe}{21}, \ion{Fe}{23}, i.e., $T\approx$ 0.4, 10, 16 MK). The 3D noise-gating technique \citep{deforest2017} was used to clean the images. We utilized SDO's Helioseismic and Magnetic Imager \citep[HMI;][]{scherrer2012} magnetograms to determine the magnetic configuration of the source region.  We employed a potential-field extrapolation code \citep{nakagawa1972} from the GX simulator package of SSWIDL \citep{nita2015}. The code was applied to a magnetogram obtained by the SDO/HMI at 17:21:45 UT on August 22, 2022.

The active region NOAA 13085 was located at N29E03 and produced multiple B- and C-class flares on August 22. This AR began emerging two days earlier, and emergence continued until around 10 UT on August 22. In addition, a small bipole started emerging inside the AR on Aug 22, during 1-10 UT. The magnetic configuration of the AR is quadrupolar (HMI magnetogram in Figure \ref{fig1}(e)). AIA 171 {\AA} images show the enlarged view of the AR (Figure \ref{fig1}(a,b)). AIA hot channel images (131, 94 {\AA}) reveal the plasma emission characteristics of a fan-spine configuration. A potential field extrapolation (Figure \ref{fig1}(f)) illustrates the distinct fan-spine topology, which aligns coherently with the plasma structure observed in AIA images. A single null resides atop the fan, and the closed outer spine is rooted in the positive-polarity sunspot P1. The width of the dome (fan separatrix) is about 30$\arcsec$. In addition, we interpret the bright feature north of the fan, visible in cool and hot AIA channels,   as a plasma sheet enveloping a preexisting current sheet (CS) (Figure \ref{fig1}(c,d)). The apparent height of the plasma sheet around the CS lies between 10-15$\arcsec$. As expected, the null point of the fan-spine configuration was embedded in the current sheet. 

The high-resolution GST images offer a detailed zoomed view of the fan-spine topology at $\approx$17:21 UT (Figure \ref{fig2}). The field of view of GST images is marked by a rectangle in coaligned HMI continuum and AIA 304 {\AA} images (\ref{fig2}(a,b)).  In Figure \ref{fig2}(c), the photospheric TiO image unveils the pores associated with the central minority polarities (positive, P2 and P2$^{\prime}$), which are surrounded by opposite polarity flux (see Fig. \ref{fig2}(a) for comparison with the HMI continuum/magnetogram) separated by a quasicircular polarity inversion line (PIL). H$\alpha$$\pm$0.8 {\AA} images reveal chromospheric loop-like structures rooted in P2$^\prime$ (Figure \ref{fig2}(d,e)). Intriguingly, the H$\alpha$ observations also reveal a vertical bright feature (marked by a white arrow) between two cusps, located around the extrapolated null point and approximately cospatial with the AIA plasma/current sheet (Fig. \ref{fig2}(b)). South and east of this feature we detected a large-scale ring of brightenings (marked by red arrows in Fig. \ref{fig2}(f)).  

\begin{figure*}[htp]
\centering{
\includegraphics[width=6.9cm]{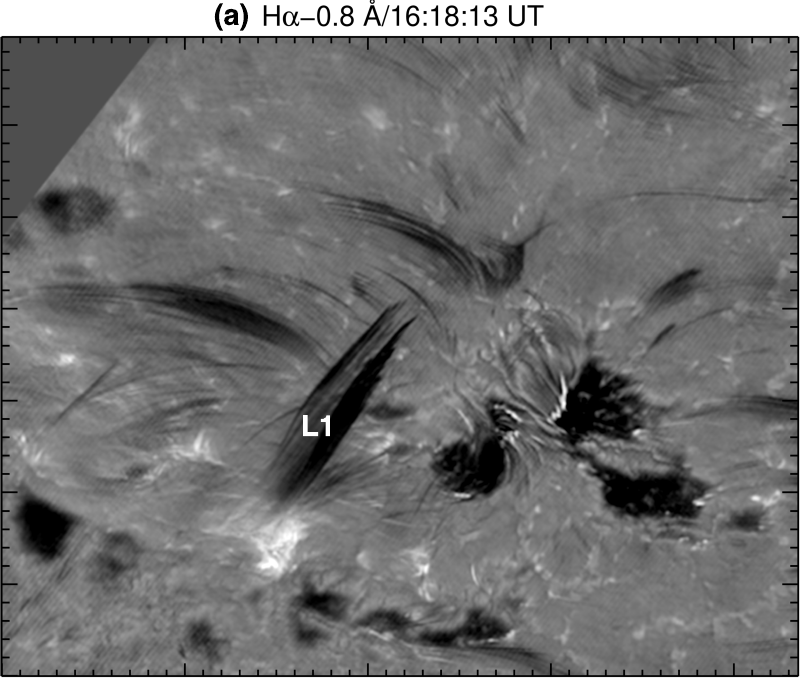}
\includegraphics[width=6.9cm]{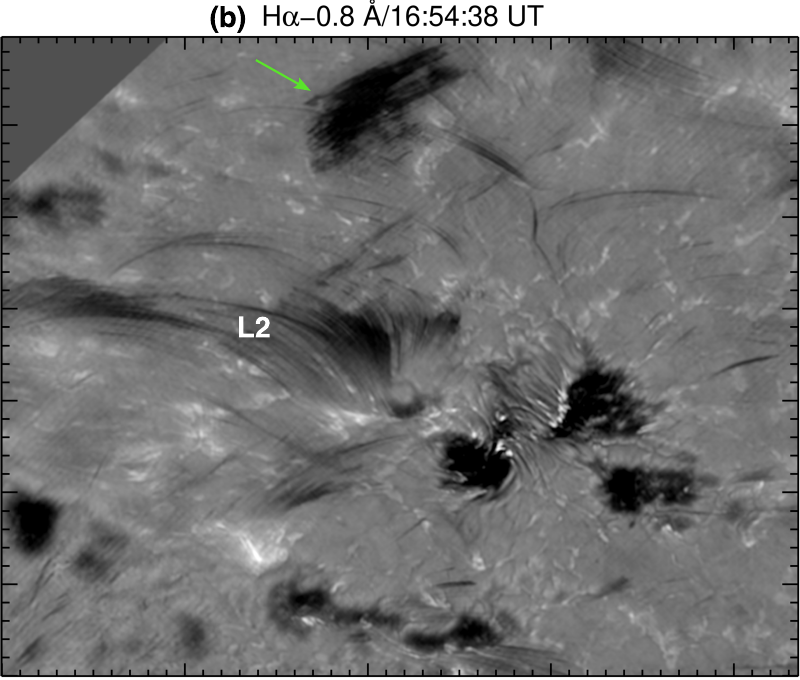}

\includegraphics[width=6.9cm]{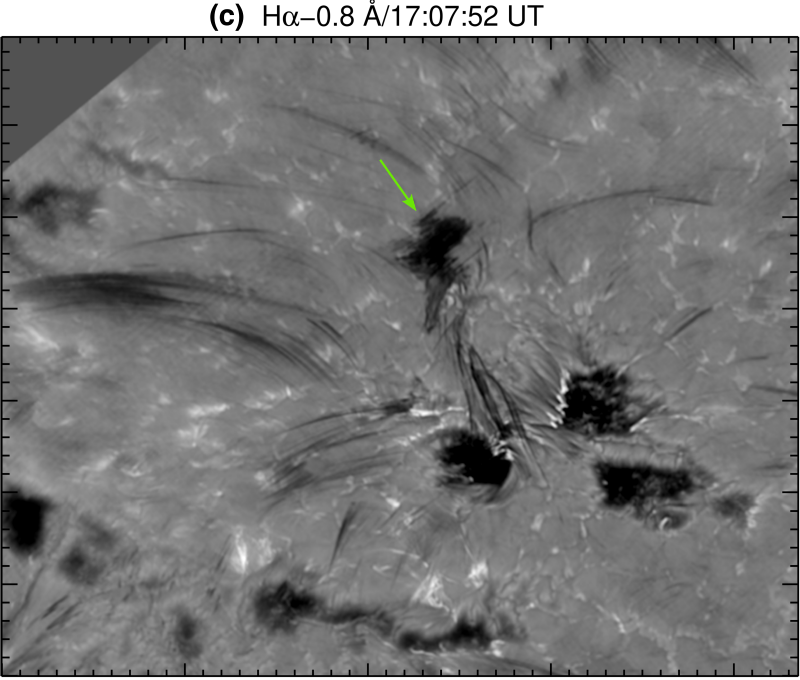}
\includegraphics[width=6.9cm]{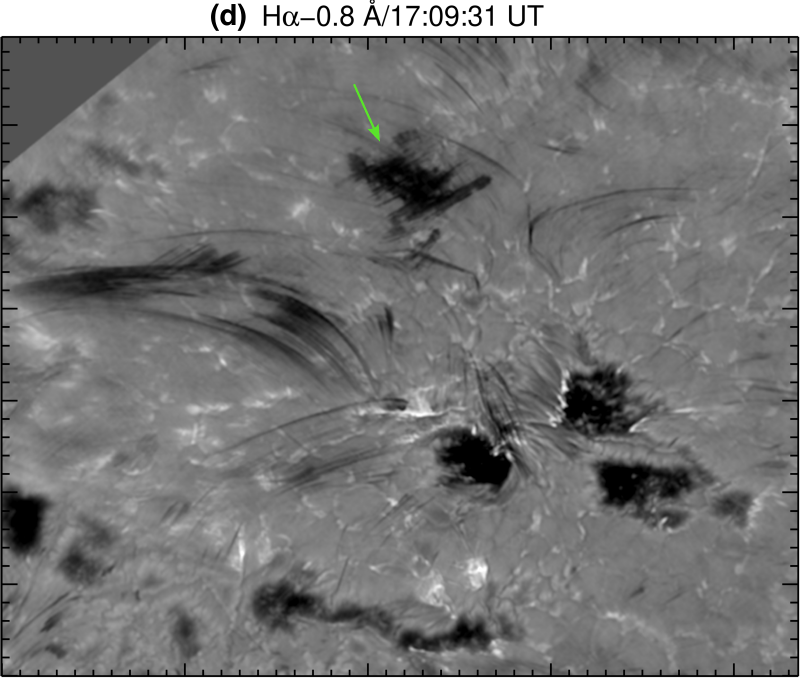}

\includegraphics[width=6.9cm]{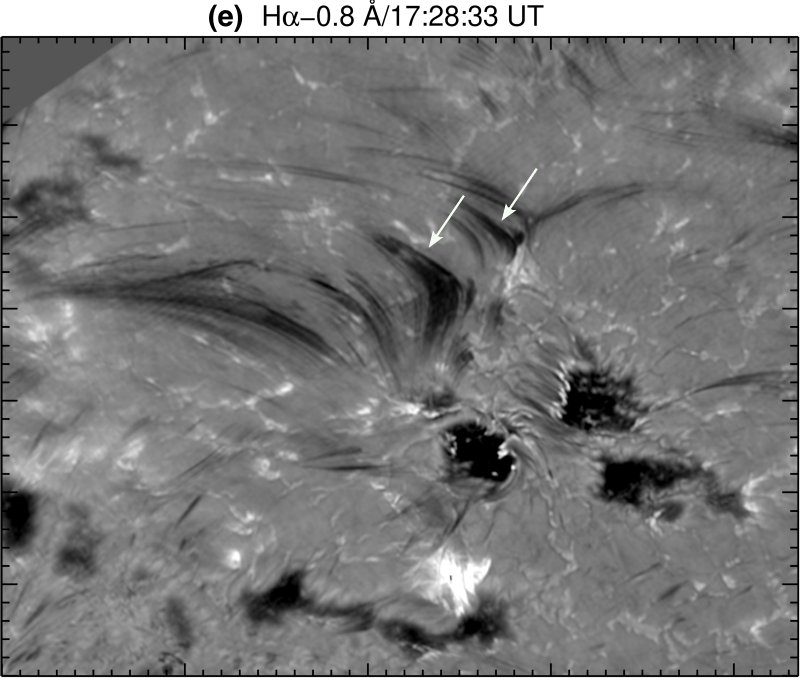}
\includegraphics[width=6.9cm]{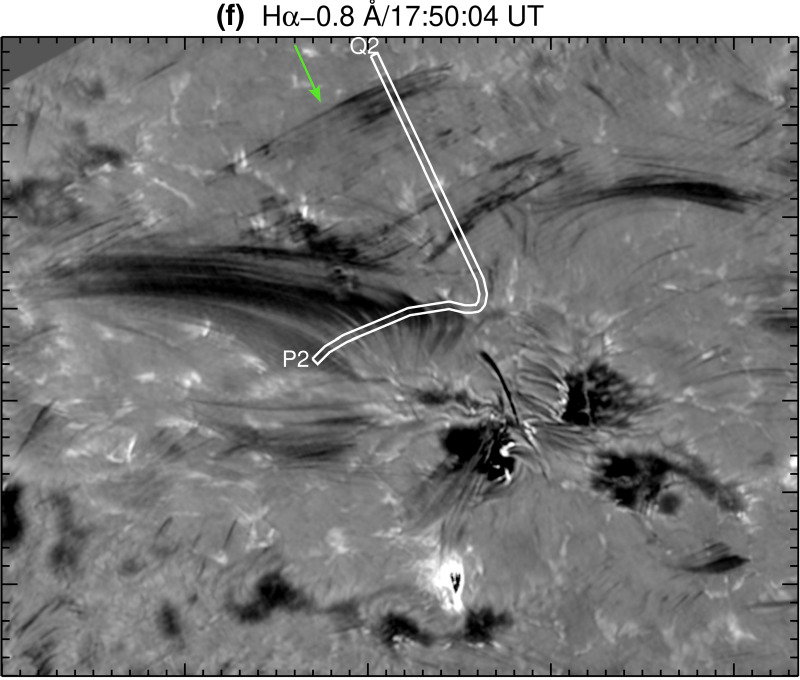}

\includegraphics[width=5.8cm]{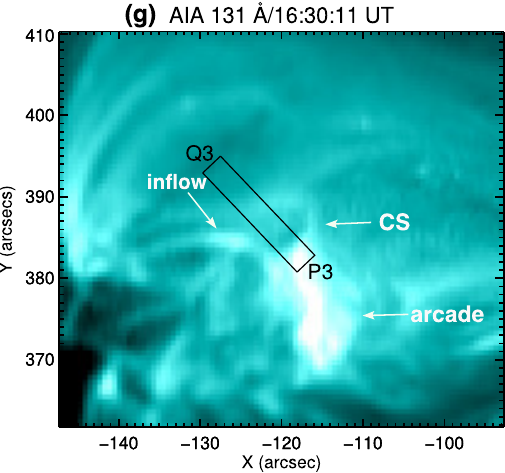}
\includegraphics[width=5.15cm]{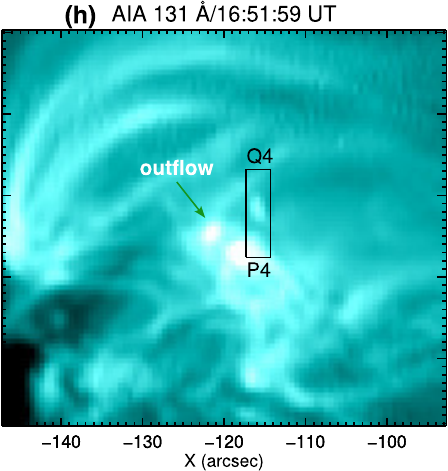}
\includegraphics[width=5.15cm]{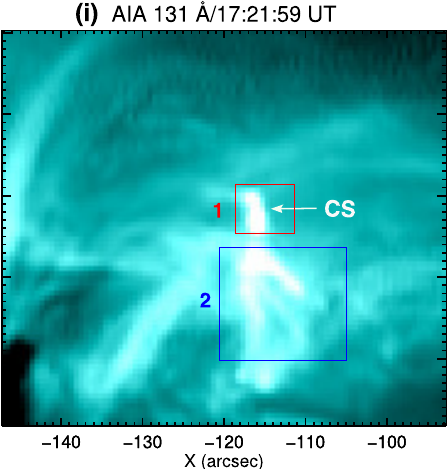}

}
\caption{{\bf Inflows and outflows.} 
(a-f) H$\alpha$-0.8 {\AA} images showing multiple loops rising from the left side of the dome; the most prominent features are labeled L1 and L2. These loops progress toward and interact with the current sheet subsequently producing outflows/jets. Green arrows = outflows, white arrows = inflows toward the CS. Each division on both axes corresponds to a distance of 1$\arcsec$. (g,h,i) AIA 131 {\AA} images display inflow/outflows in/around the bright CS and associated arcade. Slices P2Q2, P3Q3, and P4Q4 are used to construct TD intensity plots in Figure \ref{fig5}(b,c,d) respectively. Boxes 1 and 2 are utilized to extract intensity changes during magnetic reconnection.} 
\label{fig4}
\end{figure*}

The GST images in Figure \ref{fig3} further reveal a series of rising loop structures on the left (east) side of the circular PIL. These structures are best seen in H$\alpha$-0.4.  H$\alpha$-0.4 {\AA} images and the associated animation (S1.mp4) show a rising structure, denoted as L1, during 16:14-16:30 UT. Shortly thereafter, we observed another rising loop structure L2, which traveled toward the plasma sheet (Figure \ref{fig3}(b)). The western footpoint of L2 was rooted in the tiny spot P2$^\prime$ of about granular size (2-3$\arcsec$). A sequence of inflowing structures (indicated by white arrows) exemplified by L2 approaches the reconnection site and results in subsequent outflows (indicated by green arrows in Figure \ref{fig3}(b-f)). We also observed a series of features that appears to be closing down and forming an arcade above the right side of the PIL (red arrow, Figure \ref{fig3}(d)). A ring of localized brightenings was detected at the footpoints of the fan loops during this interval (Figure \ref{fig3}(e,f)). The blue-shifted chromospheric structures (e.g., L1 and L2) and the associated inflows and outflows are more clearly visible in the H$\alpha$-0.8 {\AA} images during 16:14-18:00 UT  (Figure \ref{fig4}). The accompanying animation (S2.mp4) provides a clear depiction of the series of structures that approach and retract from the plasma sheet, and the associated changes in footpoint connectivity.

AIA 131 {\AA} images and the associated animation (S3.mp4) illustrate the evolution of inflows, outflows, and associated brightenings.  As seen in the GST data, recurring inflowing loops from the left side of the dome (indicated by the arrow at 16:30:11 UT) progressed toward the bright vertical plasma sheet located above the western PIL of the fan (Figure \ref{fig4}(g)). Outflows (a bright blob marked with an arrow) were observed above the arcade (Figure \ref{fig4}(h,i)). The plasma sheet is most clearly visible at 17:21:59 UT; its dimensions are approximately 6$\arcsec$ in length and 2$\arcsec$ in width. 

\begin{figure*}
\centering{
\includegraphics[width=9.0cm]{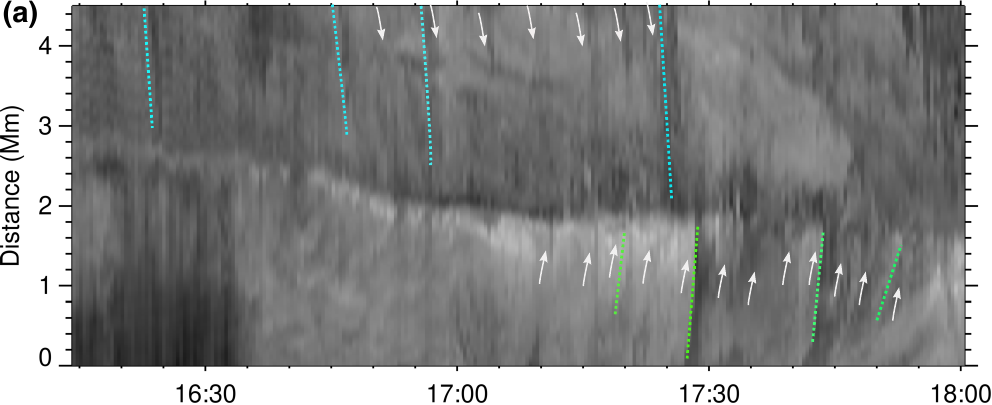}
\includegraphics[width=9.0cm]{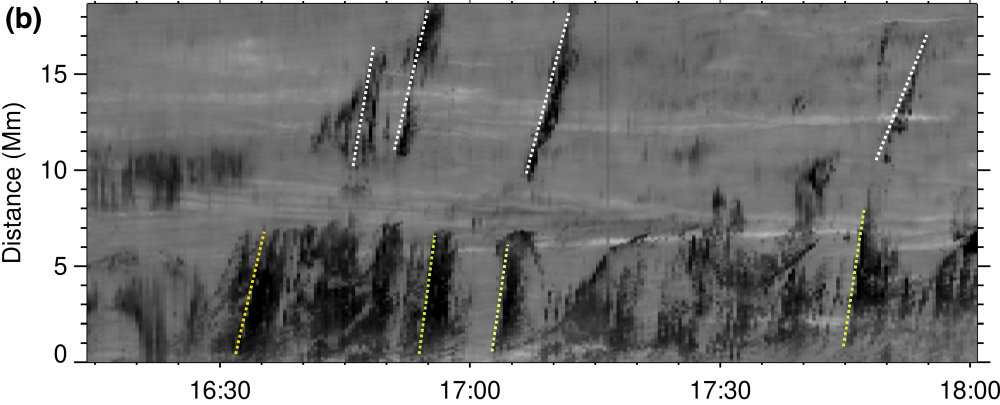}
\includegraphics[width=9.0cm]{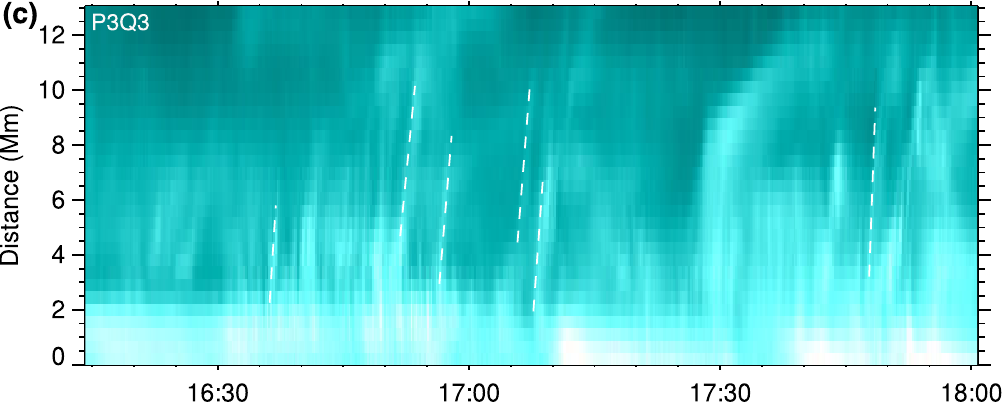}
\includegraphics[width=9.0cm]{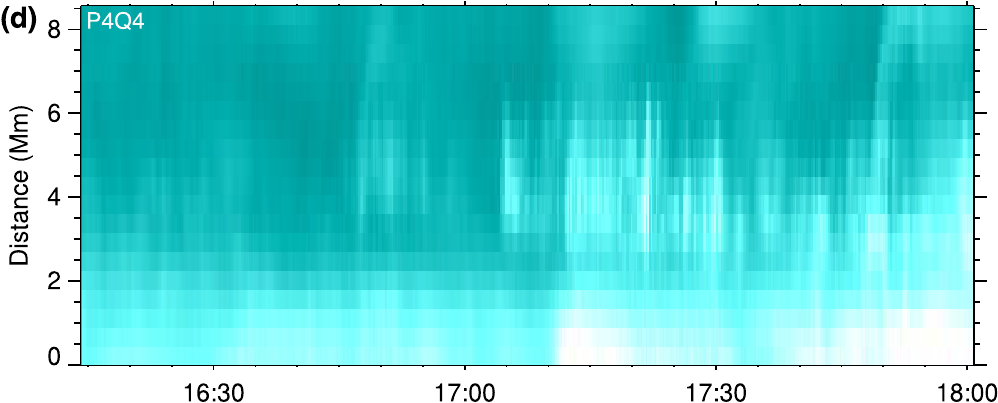}
\includegraphics[width=9.0cm]{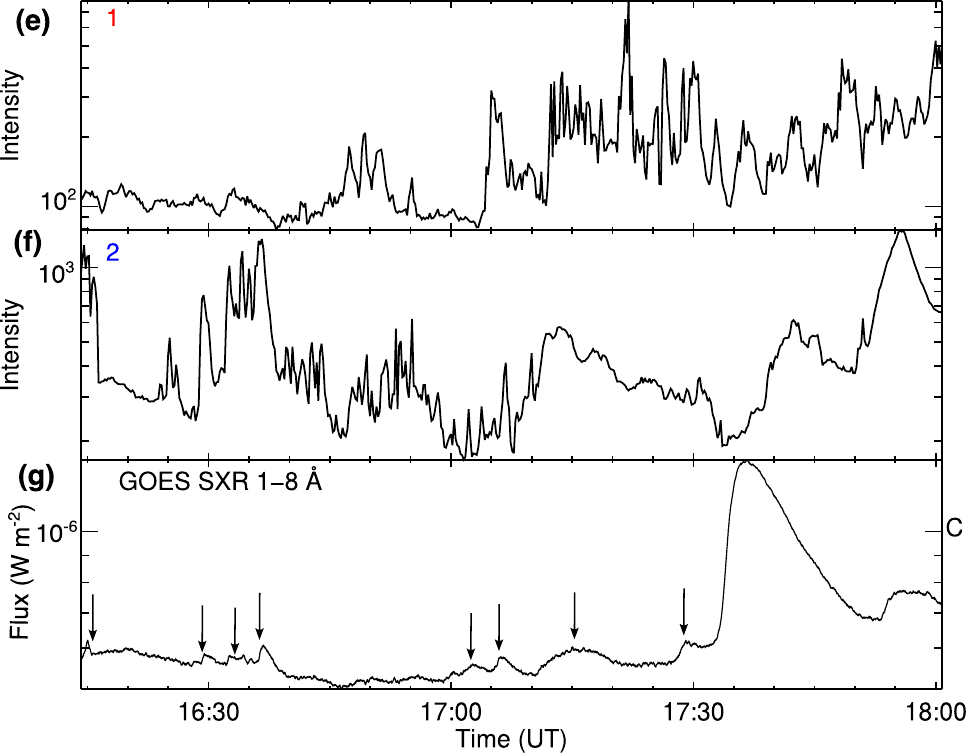}

}
\caption{{\bf Temporal evolution of inflows/outflows and EUV/X-ray intensity}. (a,b) H$\alpha$-0.4 {\AA} and H$\alpha$-0.8 {\AA} TD intensity plots along slices P1Q1 and P2Q2 in Figures \ref{fig3}(f) and \ref{fig4}(f). Some of the clearest inflows are indicated by dotted lines (cyan and green in (a)) and arrows. Lateral expansion of the rising and inflowing loops is indicated in (b) by the yellow dotted lines, while outflows are represented by the white dotted lines. (c,d) AIA 131 {\AA} TD intensity plots along slices P3Q3 and P4Q4 in Figure \ref{fig4}(g,h). Some of the outflows are indicated by the white dashed lines in (c). (e,f) AIA 131 {\AA} intensity (arbitrary unit) extracted from boxes 1, 2 in Figure \ref{fig4}(i). (g) GOES soft X-ray flux profile in 1-8 {\AA}. The arrows indicate multiple B-class flares before the onset of a C-class flare around 17:35 UT.} 
\label{fig5}
\end{figure*}

To determine the speed and frequency of inflows, we created a time-distance (TD) intensity plot along a curved slice P1Q1 (Figure \ref{fig3}(f)) across the CS to encompass inflow structures around the plasma sheet during 16:14-18:00 UT. These inflows were most clearly observed in H$\alpha$-0.4 {\AA} images. The TD intensity plot distinctly reveals recurrent inflow features moving toward the plasma sheet (indicated by white arrows in Figure \ref{fig5}(a)). We selected some of the clearest inflows for speed determination using a linear fit to the TD data points. The measured inflow speeds on the left side of the plasma sheet (green dotted lines) are 15, 20, 18, and 7 \kms, while those on the right side (cyan dotted lines) are 20, 28, 13, and 23 \kms (Figure \ref{fig5}(a)). Therefore, the estimated inflow speed ranges from 10-30 \kms.
The inflows appear to exhibit quasiperiodic behavior. The arrows indicate some of the prominent quasiperiodic features. Noting that each division on the X (time) axis represents 5 minutes, we estimate that the inflow brightness intensity varies with a period close to 3-5 minutes.

We constructed another TD intensity plot along a J-shaped slit P2Q2 (Figure \ref{fig4}(f)) to measure the recurrent motion of the inflowing loops. The lateral motion of these structures and some of the outflows are seen in the TD plot (Figure \ref{fig5}(b)).  The estimated lateral expansion speeds of the loops as they rise toward the plasma sheet are 35, 60, 37, and 48 \kms, while the outflow speeds are 46, 36, 30, and 24 \kms (Figure \ref{fig5}(b)). Note that the periodicity of the rising loop structures is approximately 5 minutes, consistent with the inflows at the plasma sheet.

The AIA 131 {\AA} TD intensity plot (Figure \ref{fig5}(c)) along P3Q3 (Figure \ref{fig4}(g)) exhibits recurrent outflows throughout the GST observations (i.e., 16:14-18:00 UT). Selected outflows (from left to right, marked by dashed lines) have speeds of 80, 60, 50, 65, 70, and 140 \kms. Additionally, the TD intensity plot along P4Q4 (Figure \ref{fig4}(h)) and the intensity profile within Box 1 (Figure \ref{fig4}(i)) reveal quasiperiodic intensity variations at the plasma sheet (Figure \ref{fig5}(d,e) and movie S4.mp4). Note that the GOES soft-X-ray (SXR) flux profile in the 1-8 {\AA} channel shows that a C-class flare peaked around 17:35 UT, after multiple B-class flares (indicated by arrows in Figure \ref{fig5}(g)). These flares are nearly simultaneous with the EUV intensity variations (Figure \ref{fig5}(f)) extracted from the flare arcade (Box 2, Figure \ref{fig4}(i)). The GOES SXR flux represents integrated emission from the entire solar disk. As there was no flare activity observed in other regions of the Sun at this time, the SXR emission primarily originates from the studied region.

The photospheric TiO images (Figure \ref{app-fig1}(a-d)) and accompanying movie (S5.mp4) reveal significant motions of the magnetic spot P2$^\prime$ (marked by an arrow in Figure \ref{app-fig1}(b-d)), where one footpoint of the rising recurrent loop structures was rooted. 
P2$^\prime$ moves from east to west and merges into the same-polarity sunspot P2 during the GST observations, 16:14-18:00 UT. The TD intensity plot (Figure \ref{app-fig1}(e)) along the slice RS shows the speed of P2$^\prime$ to be 0.85 \kms, which is typical for photospheric moving magnetic features at granular scales \citep{lamb2013}. The TiO animation (S5.mp4) shows the rotation of the P2 spot in the clockwise direction (see also the orientation change of the penumbral filament inside the green ellipse in Figure \ref{app-fig1}(b-d)). The selected HMI magnetograms (Figure \ref{app-fig1}(f-h)) display the motion and merging of P2$^\prime$, consistent with the high-resolution TiO images.

\begin{figure*}
\centering{
\includegraphics[width=4.8cm]{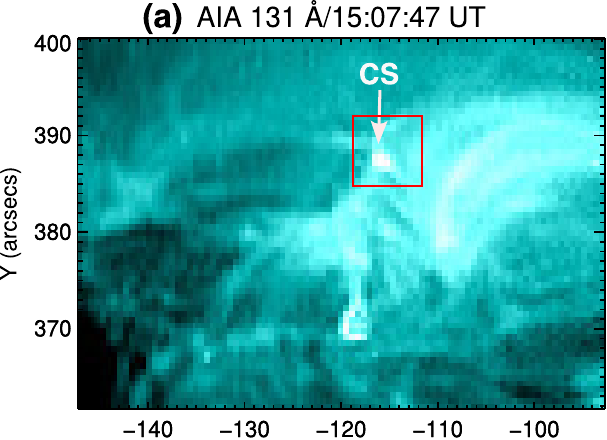}
\includegraphics[width=4.2cm]{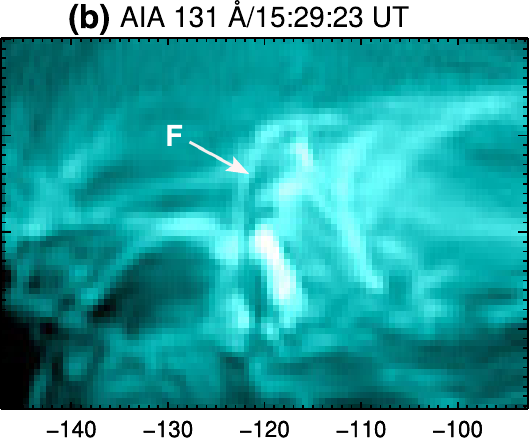}
\includegraphics[width=4.2cm]{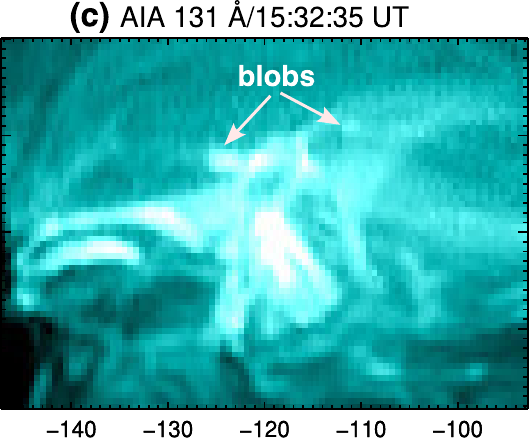}

\includegraphics[width=4.8cm]{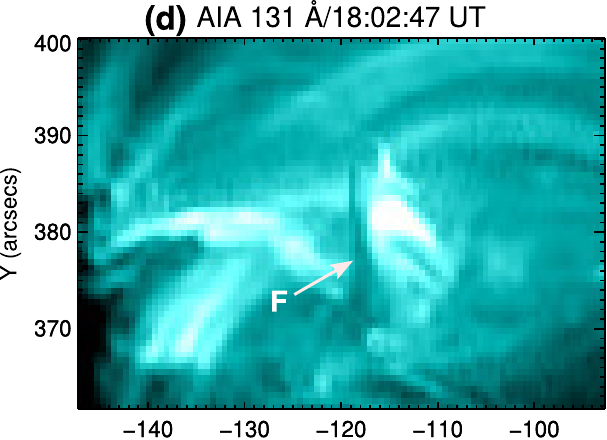}
\includegraphics[width=4.2cm]{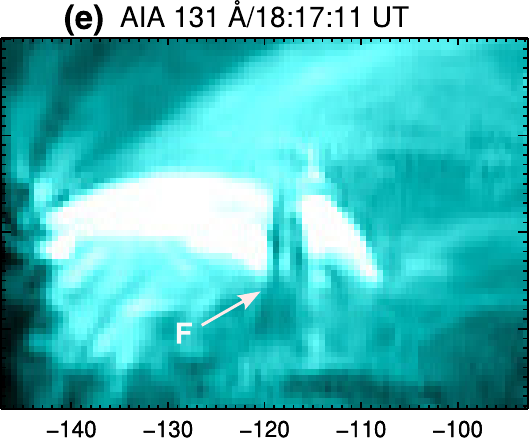}
\includegraphics[width=4.2cm]{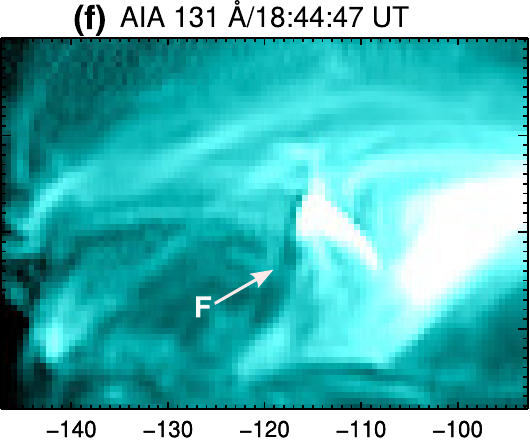}

\includegraphics[width=8.4cm]{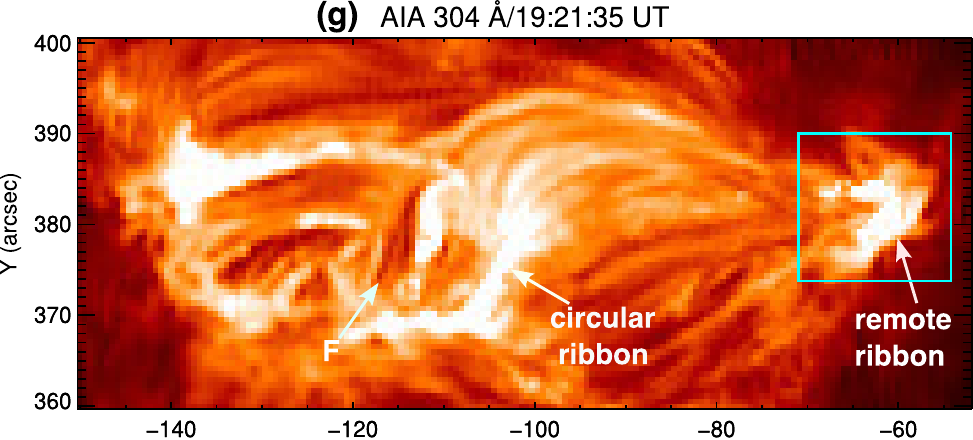}
\includegraphics[width=4.5cm]{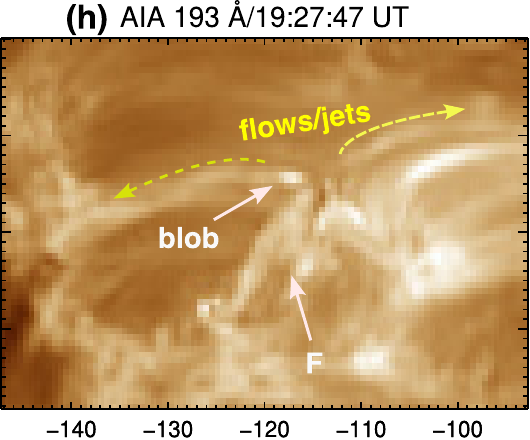}

\includegraphics[width=8.3cm]{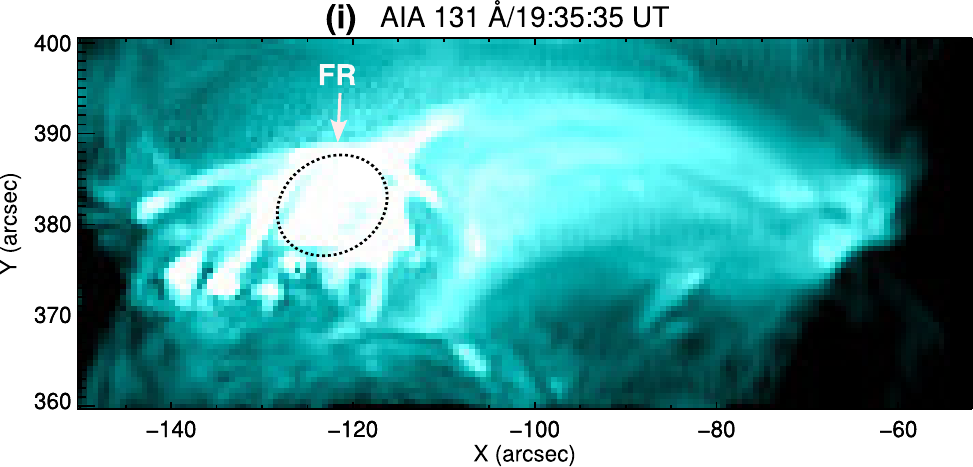}
\includegraphics[width=4.5cm]{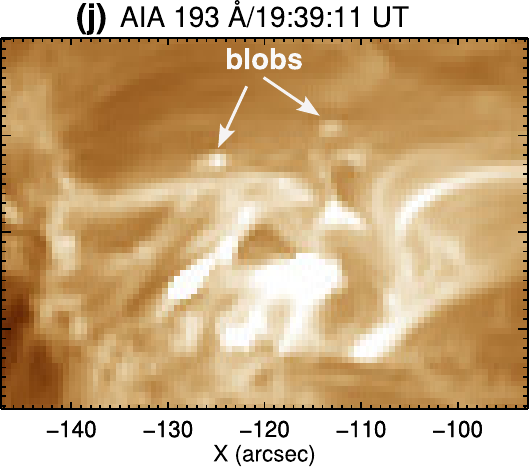}

\includegraphics[width=14.0cm]{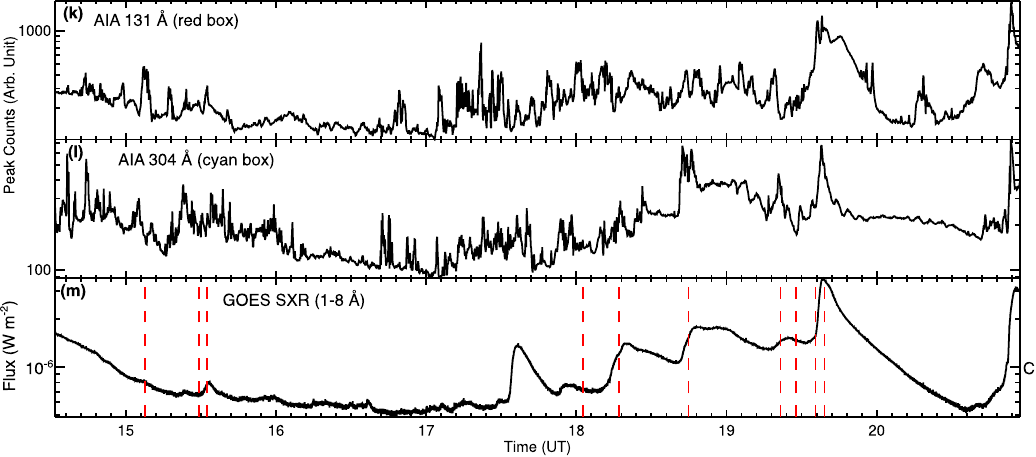}
}
\caption{{\bf Filament eruptions and associated flux rope interactions near the current sheet during B- and C-class flares}. (a-f) AIA 131 {\AA} images displaying the onset of recurrent jets and flares through filament (F) encounters at the CS. (g,i) AIA 304 {\AA} image depicting circular and remote ribbons during filament interaction with the CS. AIA 131 {\AA} image showing the encounter of the circular feature containing a filament (FR, marked by a dashed ellipse) with the CS and overlying structures during a C-class flare. (h,j) AIA 193 {\AA} images showing ejected plasma blobs from the CS. The yellow arrows indicate the bidirectional flows/jets from the CS. (k,l,m) AIA 131 and 304 {\AA} peak counts extracted from the red (panel (a) and cyan (panel (g)) boxes. Soft X-ray flux profile in the GOES 1-8 {\AA} channel. The vertical dashed lines indicate the timing of the AIA panels from (a) to (j).} 
\label{fig6}
\end{figure*}

The AIA 131 {\AA} images (Figure \ref{fig6}(a-f, i)) and accompanying animation (S6.mp4) reveal recurrent intensity fluctuations and heating at the CS from 14:31:47 to 20:57:11 UT. We observed the rising filament (marked by F) and its interactions with the CS. The filament underwent repeated activation and rising associated with the B- and C-class flares, producing multiple EUV/X-ray bursts (Figure \ref{fig6} (a-k)). 
During the B-class flares, the filament rose but did not erupt. During the C-class flares, however, the
filament on the eastern side of the PIL clearly erupted, accompanied by a quasicircular feature interacting with the CS and intense brightenings of the fan loops and loops surrounding the outer spine (Figure \ref{fig6}(i)). A circular ribbon formed simultaneously at the base of the fan and a remote ribbon appeared where the outer spine was connected to sunspot P1 (Figure \ref{fig6}(g)).

We observed multiple plasma blobs (size $\approx$ 2-3$\arcsec$) emanating from the CS during the filament eruptions (Figure \ref{fig6}(c,h,j)). 
The AIA 131 Å peak counts extracted from a box (red) containing the CS reveal multiple impulsive bursts 
(Figure \ref{fig6}(k)). AIA 304 Å peak counts extracted from the remote ribbon (cyan box in Figure \ref{fig6}(g)) show intensity fluctuations nearly correlated with the intensity fluctuations at the CS. 
The GOES soft X-ray flux in the 1-8 Å range reveals multiple B- and C-class flares throughout this active interval.  
All strong B and C-class flares coincide with the recurrent encounters between the filament and the CS. The timing of the AIA panels in Figure \ref{fig6} is marked by vertical dashed lines in the soft X-ray flux (panel (m)) to show the association between the filament eruptions and the recurrent flares.


We conducted wavelet analysis \citep{torrence1998} of the GOES SXR flux during three different intervals, spanning from 16:17 to 23:13 UT (Figure \ref{app-fig4}). Interval \#1 (16:17-17:32 UT) reveals approximate 5- and 20-minute periodicities. Both intervals \#2 (17:32-20:00 UT) and \#3 (21:25-23:13 UT) also exhibit a periodicity of about 20 minutes.
Furthermore, intensity fluctuations at the CS observed in AIA 131 {\AA} manifest periodicities of approximately 5, 10, and 20 minutes during the interval from 16:32 to 19:34 UT (bottom of Figure \ref{app-fig4}).  As noted above, all of these intervals contained recurrent B and C-class flares.

We extracted an intensity profile along slit RS from N2 using GST H$\alpha$-0.4~$\AA$ images during 16:14-18:00 UT (Figure \ref{app-fig5}(a)). The inflow structures originated from both sides of the plasma sheet (e.g., Figure \ref{fig5}(a)), and the disturbances propagating from N2 were clearest in the H$\alpha$ -0.4~$\AA$ channel. The time-distance intensity plot along RS shows outward-propagating disturbances (i.e., slow magnetoacoustic waves) (Figure \ref{app-fig5}(a)). The wavelet analysis of the average intensity signal (extracted between the two dashed lines in panel (b)) reveals periods of 3-5 and 20-30 minutes above the 95$\%$ confidence level (Figure \ref{app-fig5}(c-f)).

We performed a differential emission measure (DEM) analysis \citep{cheung2015} of the region of interest using nearly cotemporal AIA images in six EUV channels (171, 131, 94, 335, 193, 211~{\AA}) at 17:21:59 UT, when the bright plasma sheet was best visible. In the DEM code, we utilize a log T grid ranging from log T(K)=5.7 to log T(K)=7.7, with intervals of log T(K) = 0.1. The emission measures in different temperature bins are presented in Figure \ref{app-fig2}. The DEM distribution shows peaks at log T(K)=6.4, 7.1 (Figure~\ref{app-fig2} (f)). The estimated total EM, obtained by integrating the DEM distribution over the entire Gaussian temperature range, is 1.27$\times$10$^{29}$ cm$^{-5}$. Assuming that the depth of the plasma sheet along the line of sight is roughly equivalent to its width w$\approx$2$\arcsec$, the electron number density at the plasma sheet 
is $n = \sqrt{{EM}/{w}}$ 
= 2.9$\times$10$^{10}$ cm$^{-3}$ (assuming filling factor=1). 
The bright plasma sheet (CS, marked by arrows) exhibited temperatures between 10-25 MK ($\log T \mathrm{(K)}=7.0$--$7.4$) (Figure~\ref{app-fig2} (d,e)). The dome structure and hot arcade loops are most clearly observed at above 10 MK. The AIA images also show the CS in cool/warm temperatures, suggesting the presence of multithermal plasma. The estimated electron number density at the plasma sheet (on the order of 10$^{10}$ cm$^{-3}$) is consistent with previous observations (e.g., \citealt{warren2018}).

To understand the long-term evolution of the plasma sheet, we analyzed AIA 131 {\AA} images of the active region for approximately 20 hours, from 4:03:18 to 23:48:30 UT. The CS was initially detected around 4 UT. Figure \ref{app-fig3}(a,b) and the accompanying animation (S7.mp4) reveal intensity variations and the triggering of B and C-class flares for the entire 20-hour interval. The animation also displays recurrent bidirectional jets above the CS before and during the C-class flares. The GOES soft X-ray fluxes also indicate quasiperiodic fluctuations in the X-ray fluxes (Figure \ref{app-fig3}(c,d)). Figure \ref{app-fig3}(e,f) illustrates the evolution of the plasma temperature and emission measure derived from the GOES filter-ratio method \citep{garcia1994}. The plasma temperature varied between 10-14 MK, and the emission measure was about (0.05-0.2)$\times$10$^{49}$ cm$^{-3}$, in close agreement with the values derived from the DEM analysis.

 \section{DISCUSSION}\label{discussion}

 \subsection{Interpretation according to the breakout jet model}
 We analyzed high-resolution GST and AIA observations of persistent confined and eruptive activity in a small fan-spine topology on August 22, 2022. This configuration formed due to the emergence of an embedded bipole (around 1 UT on the same day) inside AR 13085. The outer spine of the topology was connected to a large positive-polarity sunspot (P1) on the right side of the AR. A fan-spine topology has a null point that is easily distorted into a current sheet by stressing of the underlying magnetic structure \citep{antiochos1998,antiochos1999}. In this case, the energy buildup was supplied by the rotational and translational motions of the sunspots, in particular the motion of P2$^\prime$ toward P2. We attribute the observed episodic energy-release signatures to bursty reconnection at the current sheet (CS) embedded in a bright plasma sheet. Repeated brightenings of the plasma sheet reflect repeated reconnection episodes, leading to outflows, local heating, and remote brightenings.

 GST high-resolution images directly reveal the reconnection site (null point) lying $\approx$15-20$\arcsec$ above the solar surface, where recurrent inflows (10-30 \kms) converged and reconnection outflows/jets (50-140 \kms) diverged. The inflow and outflow speeds measured here are consistent with previous observations of coronal inflows and outflows in large flares \citep{kumar2010,takasao2012,kumar2013,su2013,yan2022}. Although inflows were seen on both sides of the CS, the eastern (left) inflows were significantly stronger, consistent with the dominant photospheric motions on that side. The primary reconnection site barely moved during the entire 20-hour observation (Figure \ref{fig5}(a,d)) and Figure \ref{fig6}(a-f). A remarkable aspect of the observed CS is its persistence for approximately 20 hours at consistently high (10-20 MK) temperatures.  The average length (width) of the current sheet was about 6$\arcsec$ (2$\arcsec$), which implies a small aspect ratio, but this is only the apparent width and is marginally resolved by AIA. The current sheet is very likely 3D with sizeable bends and folds, so the actual width may be far smaller than what we observe. The reconnection activity at the CS continued until the decay/submergence of the embedded bipole and associated disappearance of the EUV bright point. The temporal profile of magnetic flux extracted from the red rectangular box shows a decrease in positive ($\approx$37$\%$) and negative ($\approx$37$\%$) fluxes during 15-21 UT (Figure \ref{app-fig1}(j)). The embedded bipole N2P2 decayed/submerged during multiple eruptions. However, we do not observe the formation of a flux rope at the PIL between N2P2 as would be expected in the case of flux cancellation (e.g., \citealt{amari2010,green2011,kumar2015a,kumar2017}). All eruptions originated from the multiple sheared structures at the eastern part of the circular PIL (see animation S5.mp4).
 
 The magnetic reconnection rate (i.e., ratio of inflows and outflows) was approximately 0.2, which is generally taken to be ``fast" reconnection. This also emphasizes the point that the current-sheet aspect ratio is far larger than what we infer from the observed apparent width. 
 
The breakout jet model provides a coherent framework for interpreting the observed features (Figure \ref{app-fig6}). The observed CS is a breakout current sheet, and the filament is surrounded by sheared filament-channel flux. During the B-class flares, the filament-supporting flux reached the CS and underwent some breakout reconnection, signalled by the observed intensity increases and outflows, but did not escape. This breakout reconnection removed some of the overlying field, enabling the filament-channel flux to rise, but the absence of accompanying flare signatures (e.g., a flare arcade and a bright circular ribbon) indicates that a flare current sheet didn't develop. As a result, no flux rope formed around the filament, and the breakout reconnection didn't transition to an explosive state. 

Slow reconnection in the breakout model for solar eruptions \citep{karpen2012,lynch2013,wyper2017} is a relative term, meaning it is slow in comparison to the fast reconnection during explosive breakout/flare reconnection. The slow reconnection at the breakout current sheet (BCS) removes the strapping field above the filament channel, and near-simultaneous slow reconnection below the filament builds the flux rope around the filament channel. The indirect observations of the slow reconnection are: i) pre-eruption faint jets from the breakout CS \citep{kumar2018,kumar2021} prior to the explosive reconnection during the encounter of the flux rope, and ii) multiple brightenings below the filaments during the slow rise phase.

During the larger C-class flares, however, we observed a flare arcade and the formation of a flux rope above it. The flux rope rose until it reached the breakout CS, then was partially or fully destroyed via breakout reconnection, yielding bidirectional jets/outflows and filament ejecta from the CS (animation S6.mp4). Figure \ref{fig6}(g-j) illustrates one instance of interaction between the filament-carrying flux rope (FR) and the external AR flux, leading to explosive breakout reconnection. The greater energy release characteristic of C-class flares also showed up as heating of a large portion of the fan, footpoint brightenings at the base of the separatrix (the circular ribbon) and the terminus of the outer spine (remote brightening), and the expulsion of plasmoids and bidirectional jets. The western jet was aligned with the outer spine, while the eastern jet flowed along the fan, as expected from our jet analyses and MHD simulations \citep{wyper2016}. Similar to flare ribbons, the circular ribbon and remote brightening were most likely caused by the precipitation of nonthermal electrons accelerated by reconnection at the breakout CS (Figure \ref{fig6}(g, l)). Previous observations have revealed an X-ray source located at the site of the remote ribbon in a similar fan-spine topology \citep{karpen2024}. No hard X-ray or microwave observations were available for this event.

The quasiperiodic reconnection process removed the overlying flux above the circular PIL, leading to the eruption of the filament channel. Simultaneously, flare reconnection beneath the ascending filament contributed to the formation of a flux rope around the filament. The breakout reconnection destroyed the flux rope (observable as a circular feature in AIA 131 {\AA}, Figure \ref{fig6}(i)) and generated confined bidirectional jets above the breakout CS. We observed bidirectional plasmoids above the breakout CS before and during the breakout reconnection. The size of the plasmoids ($\approx$2-3$\arcsec$) is consistent with the previous observations of plasmoids detected in breakout current sheets \citep{kumar2019a,kumar2019b}. Using the GST images (H$\alpha$+0.8~\AA) at 17:50:55 UT, \citet{cheng2024} also reported an apparent plasmoid-like feature below the plasma sheet during the decay phase of the C-class flare. During the explosive breakout reconnection, a quasicircular ribbon appeared at the base of the fan, accompanied by a remote ribbon rooted in the sunspot P1.    

Jets, plasmoids, and breakout and flare current sheets have been identified previously in individual jet events with and without mini-filaments \citep{kumar2018,kumar2019a,kumar2019b}. In those cases, the CS vanished shortly after the jet was launched. Previous MHD simulations have successfully simulated a single jet through breakout reconnection associated with the eruption of a filament channel, demonstrating the viability of the breakout jet model \citep{wyper2017,wyper2018}. The resistive-kink model also produces repetitive jets through breakout reconnection as long as the broad rotational footpoint shearing is continued \citep{pariat2009,pariat2010}. However, in the current study, the CS was detected for many hours,  reappearing again and again after being obscured by the explosive reconnection associated with the C-class flares and their filament eruptions. The GST observations clearly show a series of loop-like structures that rose toward the CS during the converging motion/rotation of the embedded bipole (N2P2 and P2$^\prime$) at the base of the fan-spine topology. Despite the long-term reconnection process transferring substantial flux from the east side of the fan to the west, the fan separatrix (footpoints marked by the circular ribbon) did not move significantly to the west. The strong sunspot field on that side undoubtedly prevented expansion of the fan and associated relocation of the separatrix.
No MHD simulations thus far have depicted such a long-lived CS undergoing repeated episodes of magnetic reconnection, but no simulations have implemented the long-term rotation and converging motion of the underlying photospheric field observed in the present case. This scenario is ripe for further numerical study.

The recurring filament eruptions are confined, most likely because the source region is located in the center of the AR, which has a significant/strong overlying magnetic flux rooted in the parent sunspot pair (N1P1). In addition, our prior studies have shown that only jets are produced when the flux rope is mostly destroyed via breakout reconnection at the null, leaving insufficient flux in the rope to erupt. Overall, the observations are consistent with the prediction of the universal breakout model for solar eruptions/jets \citep{wyper2017,wyper2018}.
\subsection{Quasiperiodic pulsations}
The findings outlined in this study carry significant implications for understanding the Quasiperiodic Pulsations (QPP: \citealt{nakariakov2009,doorsselaere2016,mcLaughlin2018,zimovets2021}) phenomenon in solar flares and jets. Our observations of recurrent reconnection unveiled QPP in both EUV and X-ray intensity, as well as in the structured inflows seen by GST in H$\alpha$. The wavelet analysis and TD intensity plots find distinct periodicities of approximately 5, 10, and 20 minutes in the X-ray and EUV peak intensities at the CS, and 5 min for the inflowing cool structures. 

We attribute the 5-minute QPPs observed during the small B-class flares to repetitive bursts of reconnection and resulting jets and heating. In the case of the C-class flares, the X-ray light curve displays a distinctive pattern of growing and decaying QPPs, characterized by a period of approximately 20 minutes (Figure \ref{app-fig4}). Simultaneous AIA observations reveal recurrent activation and eruption of filaments at the same intervals. Therefore the longer-period QPPs are associated with filament eruptions.  

What triggers the recurrent inflows towards the CS? The inflows are clearly evident at chromospheric levels in the very high-resolution blue-shifted emission observed by GST, and are inferred from the lower-resolution AIA data. The observed rotation and translational motions of the embedded bipole N2P2 and P2$^\prime$ are the most likely driver for the east-side flows themselves, but do not explain the periodicities.  
 The leakage of upward-moving slow-mode waves (p-mode with 3-5 minute periods) from P2, P2$^\prime$, and N2 may play a role in creating denser chromospheric structures at regular intervals and pushing them toward the plasma sheet. The leakage of p-mode waves also has been proposed as a trigger for chromospheric/transition region explosive events \citep{chen2006}, as supported by both MHD simulations \citep{heggland2009} and observations of periodic emissions from fan-spine topologies in active regions and coronal holes \citep{kumar2015,kumar2016,kumar2022}. An alternative explanation for the repetitive nature of the activity is that sufficient free energy needs to build up between reconnection episodes to thin down the current sheet and power the next eruption. It is unclear, however, why 5 or 20 min would be required in this case.

Similar periodicities (5, 10, 20 minutes) have been identified in small null-point topologies and jetlets observed at the base of coronal plumes, and in the switchbacks detected with Parker Solar Probe (PSP) during Encounter 10 \citep{kumar2022, kumar2023}. However, resolving the reconnection sites in tiny fan-spine topologies using AIA observations proved challenging. The slow magnetoacoustic waves emanating from N2 exhibit periods ranging between 3-5 minutes and 20-30 minutes. Sunspot observations have revealed chromospheric shocks occurring at 20-min intervals \citep{yurchyshyn2015}, which could provide added repetitive ``kicks" for the observed filament eruptions. The observations reported here provide direct insights into the 5, 10, and 20-minute periodicities in fan-spine topologies. Note that here we report jets in a fan-spine topology with an outer spine that closes in the active region, resulting in all mass and energy being confined within the AR. In contrast, for fan-spine topologies inside coronal holes and at the base of plumes, the outer spine is open and conveys mass and energy into the solar wind, potentially initiating microstream/switchbacks \citep{kumar2022,wyper2022,raouafi2023,uritsky2023}.

The sequence of small B-class flares prior to the C-class flares also could be considered as very long period pulsations (preflare VLPs, period=1.9-47.3 minutes), similar to those reported by \citet{tan2016}. They suggested that these VLPs are the manifestation of LRC oscillations of current carrying loops or MHD oscillations of coronal loops. According to \citet{zimovets2022}, however, these VLPs may be due to episodic energy release via magnetic reconnection. Our observations support the latter scenario as a source of VLPs.

The solar QPPs detected in X-ray and EUV intensities play a crucial role in understanding the origin of QPPs on other stars. Here we directly capture the recurrent null-point reconnection associated with filament eruptions as a source of QPPs. This mechanism is likely to occur during flares on other Sun-like stars as well, and could account for both growing (amplitude increasing with time) and decaying (amplitude decreasing with time) QPPs during stellar flares \citep{anfinogentov2013,cho2016,pugh2016,broomhall2019}.

 \subsection{Impulsive heating}
 Our observations provide evidence for the impulsive heating of active regions through recurrent reconnection, specifically via the DC (Direct Current) heating mechanism. The footpoint motion of the magnetic structures rooted in P2$^\prime$ and the rotation of P2 increases the free magnetic energy of the coronal field. The Poynting flux due to the footpoint motion using the formula F=-$\frac{1}{4\pi}$B$_{v}${\bf B$_{h}$}$\cdot$ {\bf V$_{h}$}, where B$_{v}$ and {\bf B$_{h}$} are horizontal and vertical components of magnetic field \citep{klimchuk2006}. {\bf V$_{h}$} is the horizontal velocity of the footpoint. From our observations, B$_{v}$$\approx$300 G (Figure \ref{app-fig1}). {\bf V$_{h}$}=0.85 \kms. Assuming B$_{h}$=B$_{v}$. The estimated Poynting flux is 1.65$\times$10$^8$ ergs cm$^{-2}$ s$^{-1}$, which is about one order of magnitude larger than the required coronal energy losses (10$^7$ ergs cm$^{-2}$ s$^{-1}$) in active regions \citep{withbroe1977,sakurai2017}. 

Are these quasiperiodic reconnection jets and associated small flares enough to heat the active region? We estimate the thermal energy of the flares using E=3nkTV=$4.14\times 10^{-16}T\sqrt{EM.V}$, where T is the plasma temperature, V is the volume, and EM is the emission measure. We assume a plasma filling factor=1. During reconnection and jetting, fan loops and reconnected loops rooted in P1 were heated beyond 10 MK and likely served as conduits for accelerated particles. Consequently, quasicircular and remote ribbons formed during the recurrent reconnection at the CS (Figure \ref{fig6}(g)). The total volume of the heated region is approximately $\frac{2}{3}\pi r_1^{3}+\pi r_2^{2}L$, assuming spherical and cylindrical shapes for the dome and outer spine structures. Here, L is the length of the outer spine (45$\arcsec$), the dome width is about 30$\arcsec$ (with a radius $r_1$ of 15$\arcsec$), and the width of the outer spine loops is about 6$\arcsec$ (with a radius $r_2$ of 3$\arcsec$, Figure \ref{fig6}(i)). The estimated volume is 3.5$\times$10$^{27}$ cm$^3$.

We observed B-class flares with shorter periods (5 min) and C-class flares with longer periods (20 min). The average emission-measure values during the B- and C-class flares are $0.03\times10^{49}$ and $0.1\times10^{49}$ cm$^{-3}$ respectively (Figure \ref{app-fig3}(c)). Utilizing the above emission measures, volume, and an average temperature of 12 MK (Figure \ref{app-fig3}(c)), the estimated thermal energies are $1.5\times10^{29}$ and $3.0\times10^{29}$ ergs. The average thermal energy of these B- and C-class flares is roughly one order of magnitude higher than the upper limit of the energy range for microflares (10$^{26}$-10$^{28}$ ergs). 
The area of the AR covering the fan-spine topology is approximately $90\arcsec\times40\arcsec$ (Figure \ref{fig1}). Assuming the periods quoted above, therefore, the flares and associated outflows occur at a rate of $1.6\times10^{-22}$ cm$^{-2}$ s$^{-1}$ for B-class flares and $4.1\times10^{-23}$ cm$^{-2}$ s$^{-1}$ for C-class flares. The typical thermal energy flux is $2.4\times10^7$ ergs cm$^{-2}$ s$^{-1}$  for B-class flares and $1.2\times10^7$ ergs cm$^{-2}$ s$^{-1}$ for C-class flares. These values are comparable to the energy flux required to heat the magnetically connected portions of the active region (10$^7$ ergs cm$^{-2}$ s$^{-1}$).

This study will help in developing MHD models of coronal heating in active regions by highlighting the potential role of null-point reconnection in the impulsive heating processes within active regions. 
The recurrent nature of the observed reconnection events implies the existence of underlying periodic drivers in terms of filament eruptions and p-mode waves, prompting further investigation into the triggers and mechanisms governing this periodic behavior. Understanding these mechanisms is crucial for refining theoretical frameworks and advancing our ability to predict and model the dynamic behavior of solar active regions.

 \section{CONCLUSIONS}\label{conclusions}
We have demonstrated direct imaging of quasiperiodic magnetic reconnection, including inflows and outflows, occurring at a current sheet near a null point in the fan-spine topology. These observations, with an ultra-high spatial resolution of approximately 50 km, provide an unprecedented view of the surrounding prolonged plasma sheet's structure. Furthermore, this study marks the first reported detection of a prolonged plasma/current sheet in the fan-spine topology and associated quasiperiodic magnetic reconnection that lasted about 20 hours. 
The existence of the breakout CS and pre-eruption magnetic reconnection there, punctuated by intermittent eruptive flares, is consistent with the breakout model for solar jets \citep{wyper2017}. Pre-eruption reconnection removes the overlying flux above the filament channel, while slow reconnection in the underlying flare CS builds up a rising flux rope. The C-class eruptive flares occur when the flux rope encounters the breakout CS, causing feedback between the explosive breakout and flare reconnections and driving an associated jet.  MHD simulations are needed to fully understand the existence of such a prolonged current sheet, the driver for the quasiperiodic reconnection, the 20-min interval between eruptions, and the impact of the associated impulsive heating on the magnetically connected corona.
These observations have implications for understanding the initiation of coronal jets, quasiperiodic pulsations, and the impulsive heating of the active region through recurrent precipitation of electrons to the base of the fan and the remote footpoint connected with the outer spine. Further observational studies are needed to establish whether this heating method is prevalent in other active regions.

The high-resolution observations reported here will inspire new MHD simulations and laboratory plasma experiments for sustained quasiperiodic reconnection as a source of recurrent jets and plasma heating. For example, MHD simulations should be conducted to understand the nature of quasiperiodic reconnection at the null-point current sheet formed within small fan-spine topologies, which are ubiquitous on the Sun. Furthermore, the role of p-mode waves in triggering or modulating recurrent null-point reconnection requires further exploration through high Lundquist number computational experiments. The combination of repetitive filament-channel eruptions with p-mode driving is another intriguing direction for numerical simulations. This study also contributes significant insights into the origin of X-ray QPPs in Sun-like stars. By directly capturing the recurrent  reconnection associated with filament eruptions as a source of QPPs, we broaden our understanding of these phenomena, not only within the solar context but also in other stars.  
In the future, high-resolution observations by DKIST (Daniel K. Inouye Solar Telescope: \citealt{rast2021}) and advanced space missions such as MUSE (Multi-slit Solar Explorer; \citealt{dePontieu2022}) will probe similar prolonged current-sheet events, reconnection jets, and associated plasma heating. \\
\\

\noindent
{\bf {Acknowledgements}}\\
We are grateful to the reviewers for their constructive comments/suggestions, which have improved the paper. SDO is a mission for NASA's Living With a Star (LWS) program. 
This research was supported by NASA's Heliophysics Guest Investigator (\#80NSSC20K0265), supporting research (\#80NSSC24K0264), LWS (\#80NSSC22K0892), Internal Scientist Funding Model (H-ISFM) programs, and NSF SHINE grant (\#2229336). 
We gratefully acknowledge the use of data from the Goode Solar Telescope (GST) of the Big Bear Solar Observatory (BBSO). BBSO operation is supported by US NSF AGS-2309939 and AGS-1821294 grants and New Jersey Institute of Technology. GST operation is partly supported by the Korea Astronomy and Space Science Institute and the Seoul National University. V.Y. acknowledges support from NSF AGS 2309939, 2300341, and AST 2108235 grants. Wavelet software was provided by C. Torrence and G. Compo, and is available at \url{http://paos.colorado.edu/research/wavelets/}.
Magnetic-field extrapolation was visualized with VAPOR (\href{(www.vapor.ucar.edu}{www.vapor.ucar.edu}), a product of the Computational Information Systems Laboratory at the National Center for Atmospheric Research.\\
\\

\bibliographystyle{aasjournal}
\bibliography{reference.bib}



\appendix
\counterwithin{figure}{section}
\section{Footpoint converging motion and merging}
\begin{figure}[htp]
\centering{
\includegraphics[width=7.8cm]{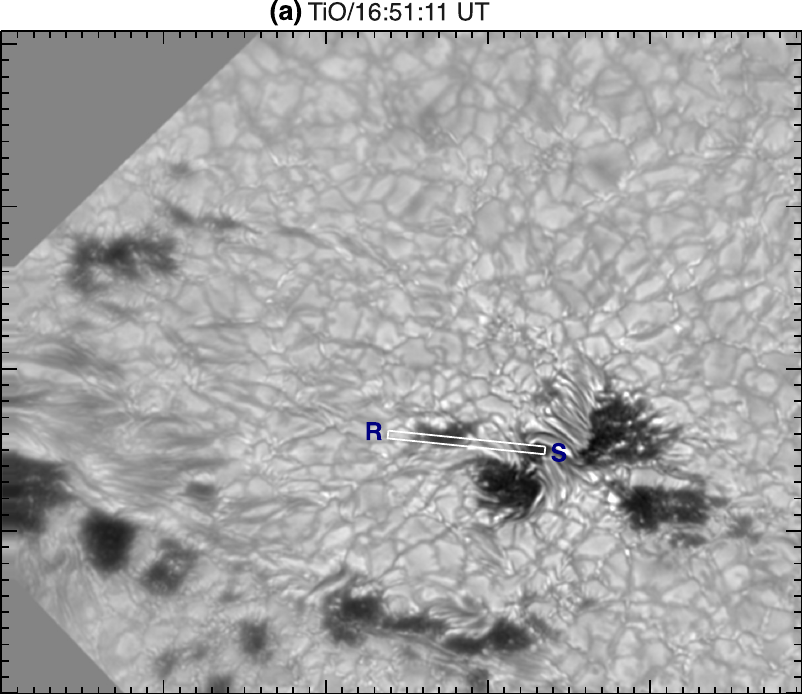}
\includegraphics[width=9.6cm]{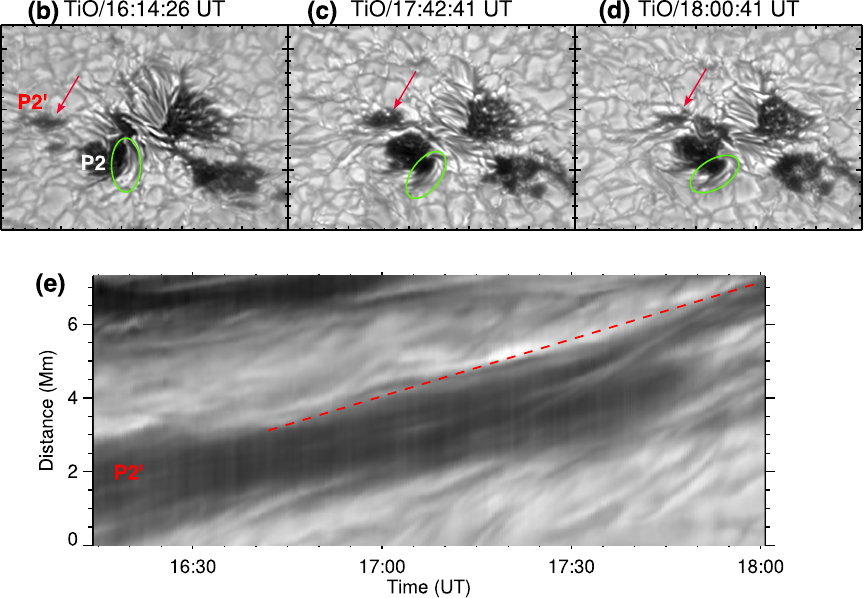}

\includegraphics[width=6.0cm]{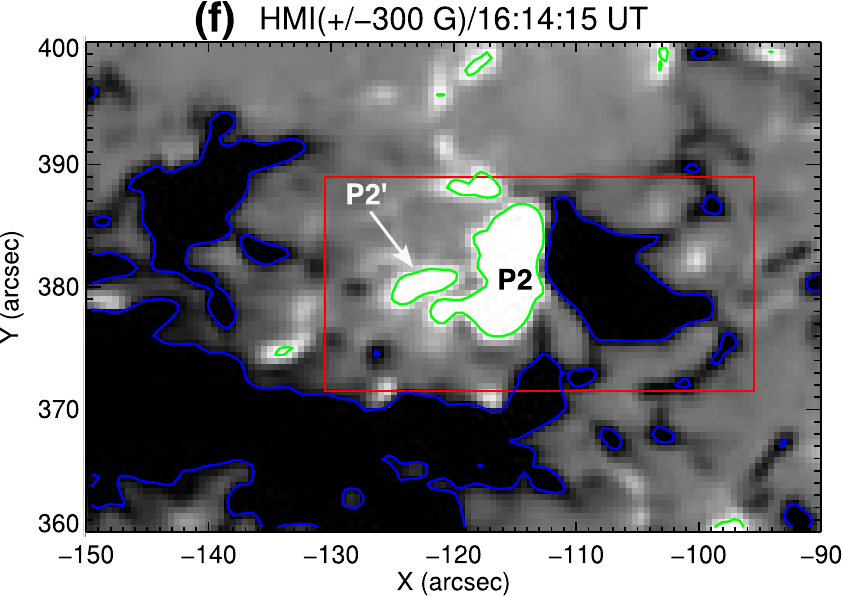}
\includegraphics[width=5.6cm]{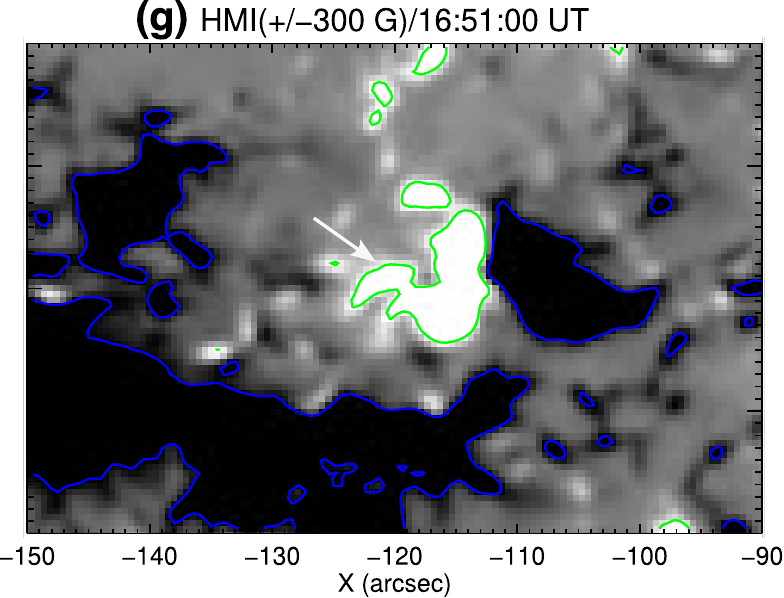}
\includegraphics[width=5.6cm]{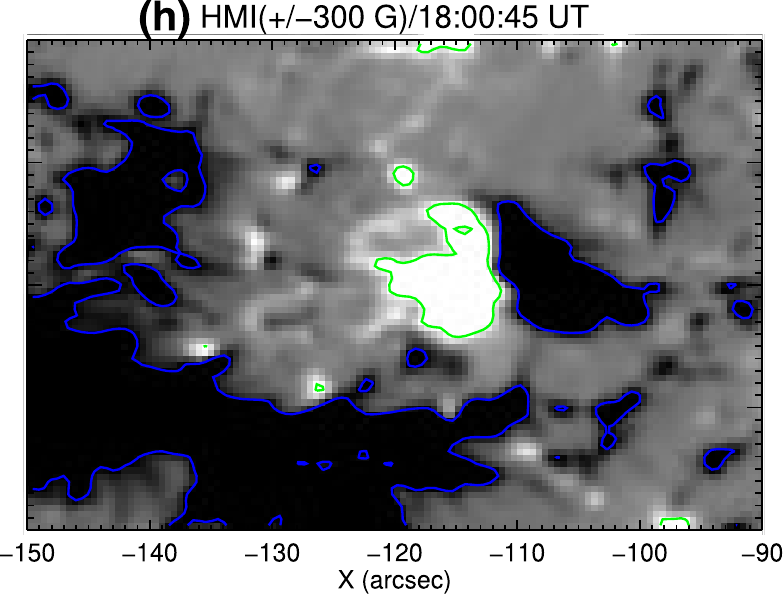}
\includegraphics[width=14.8cm]{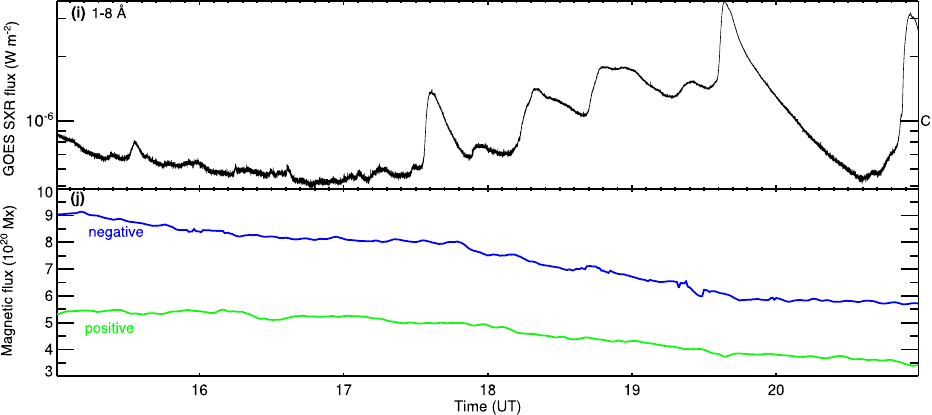}
}
\caption{{\bf Evolution of the photospheric magnetic field at the base of the fan-spine flux system.} (a-d) Photospheric TiO images depict the converging motion and merging of P2$^\prime$ (indicated by red arrows) into P2 during 16:14-18:00 UT. The green ellipse illustrates the clockwise rotation of the penumbral filament. (e) TiO time-distance intensity plot along slice RS. The translational speed of P2$^\prime$ is 0.85 \kms. Panels (f), (g), and (h) are selected HMI magnetograms (scaled between $\pm$300 G) showing the evolution of the photospheric magnetic field. The green (blue) contours (levels=$\pm$300 G) indicate positive (negative) polarities. The arrows show that P2$^\prime$ merges into P2. (i) GOES SXR flux in the 1-8 $\AA$ channel. (j) Positive/negative (green/blue) flux profiles extracted within the red box in panel (f) during 15:00-21:00 UT.} 
\label{app-fig1}
\end{figure}
\clearpage


\section{DEM analysis}
\begin{figure}[htp]
\centering{
\includegraphics[width=13.0cm]{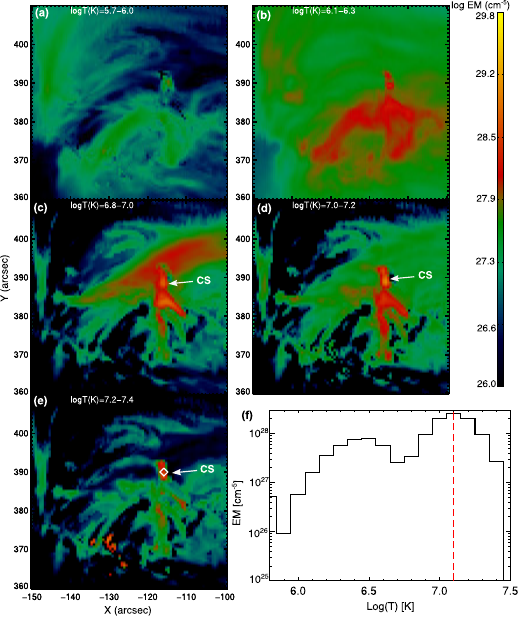}
}
\caption{{\bf (a-e) DEM maps of the CS and surroundings at various temperatures} derived using near-simultaneous six-channel AIA images at 17:21:59 UT. The arrows indicate the bright plasma sheet. The color coding indicates the total EM within the log T range marked in each panel. (f) EM profile of the plasma sheet (marked by a white diamond in panel (f)). The vertical dashed lines indicate the EM peak at log T(K)=7.1.}  
\label{app-fig2}
\end{figure}
\clearpage
\section{Long term evolution}

\begin{figure}[htp]
\centering{
\includegraphics[width=13.0cm]{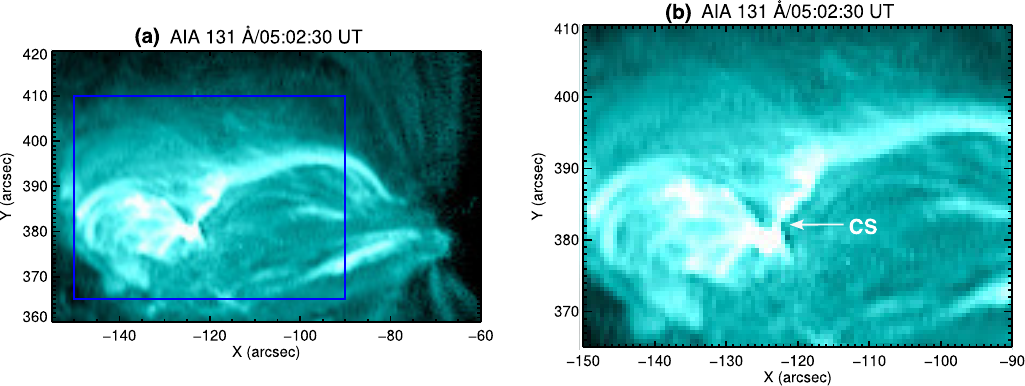}
\includegraphics[width=13.0cm]{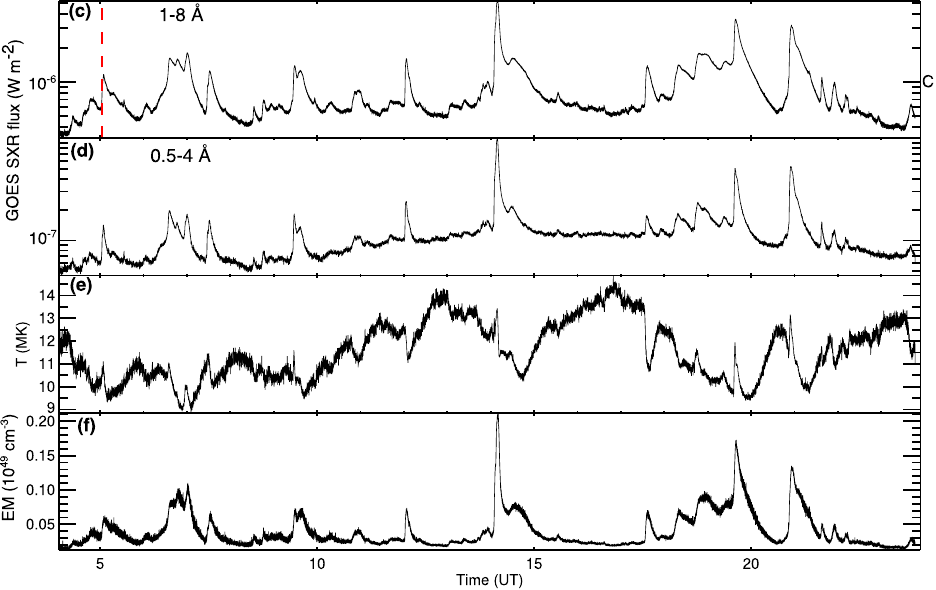}
}
\caption{{\bf Long-term evolution (20 hours) of the fan-spine topology.} (a,b) AIA 131 {\AA} images during a C-class flare (marked by a dashed line in (c)). Note that several C-class flares occurred during the entire interval. (c,d) Soft X-ray fluxes in GOES 1-8 and 0.5-4 {\AA} channels. (e,f) Temperature and emission measure derived from the GOES filter-ratio method.} 
\label{app-fig3}
\end{figure}

\clearpage

\section {Wavelet analysis}
\begin{figure}[htp]
\centering{
\includegraphics[width=8.0cm]{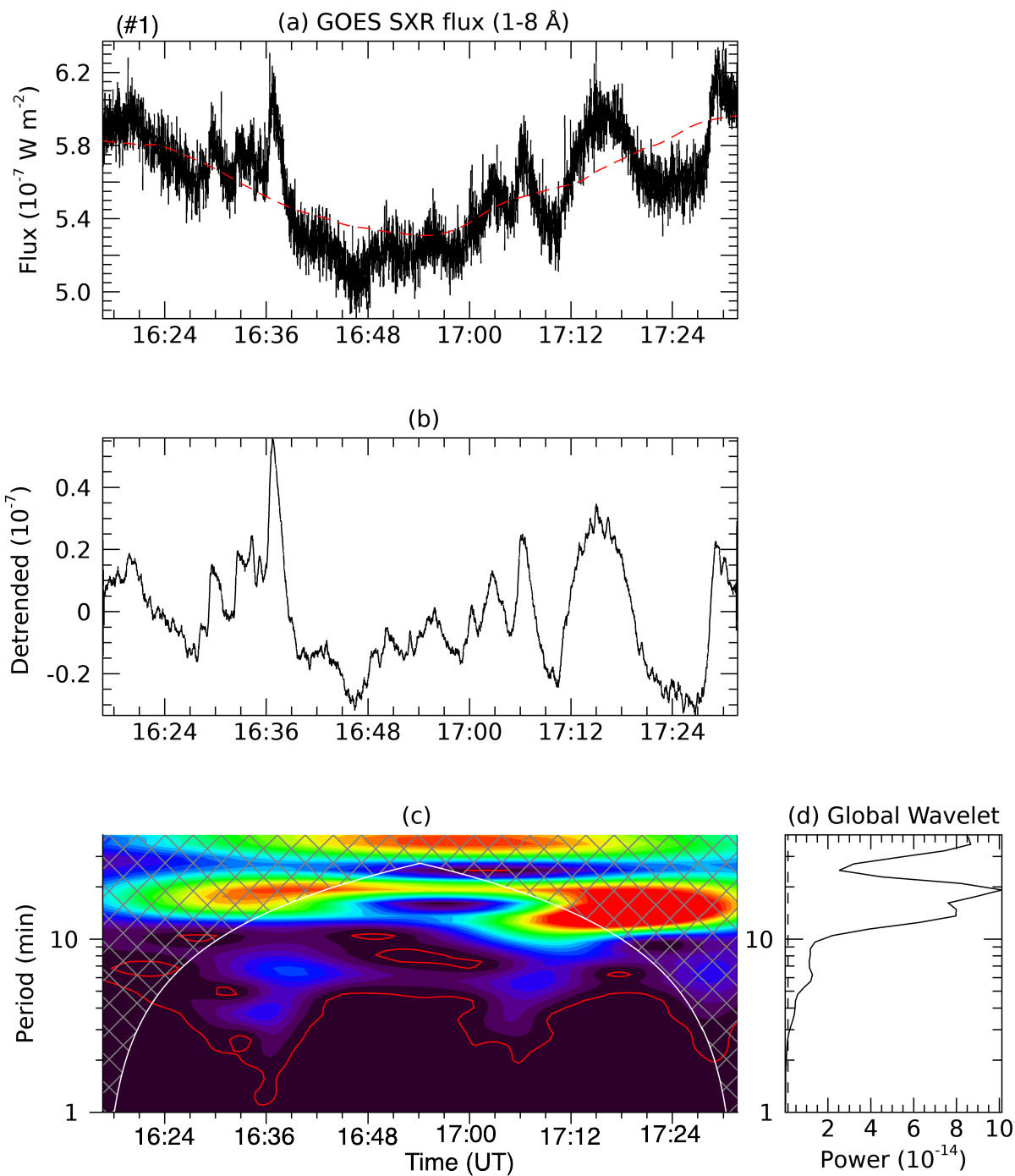}
\includegraphics[width=8.0cm]{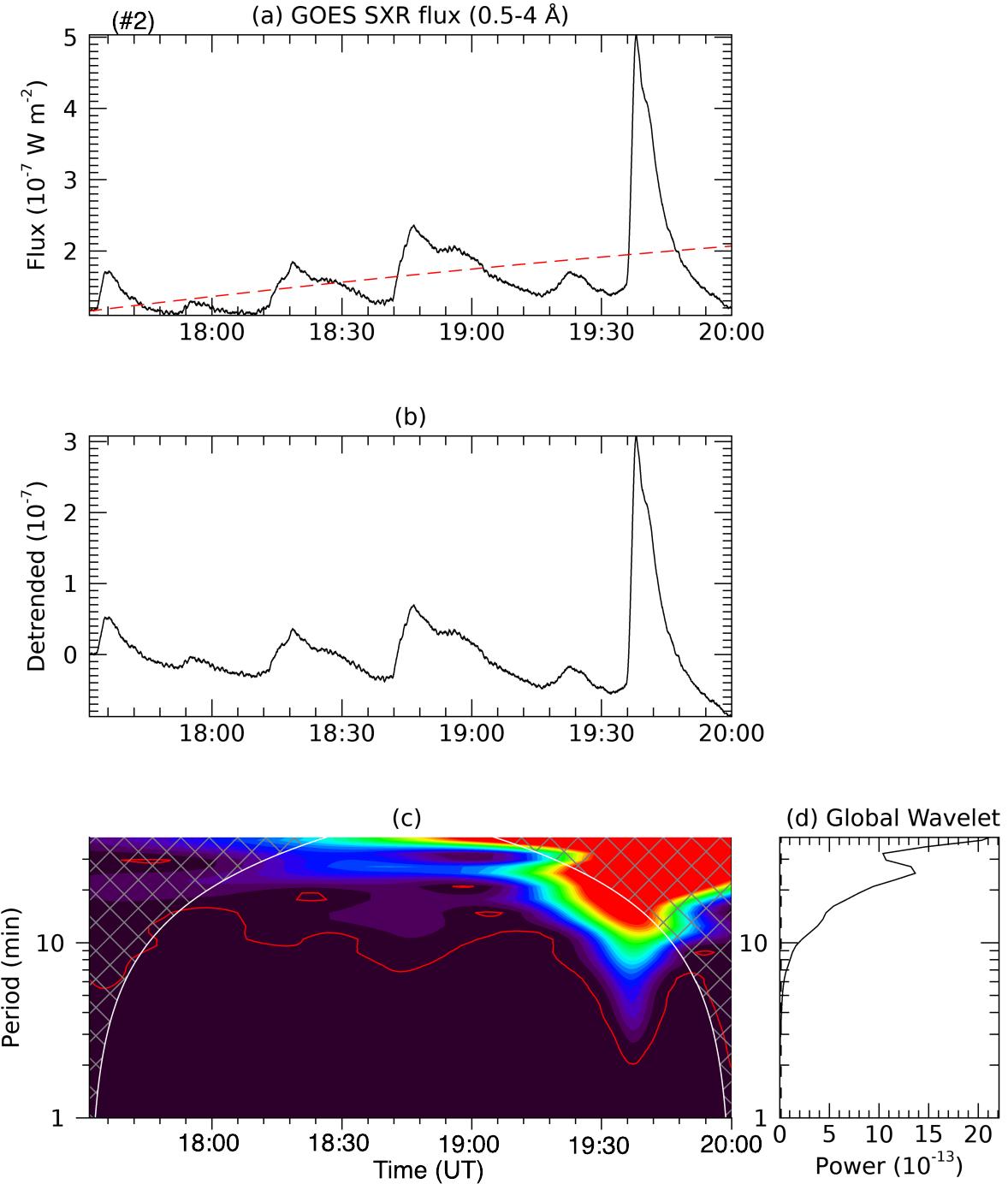}

\includegraphics[width=8.0cm]{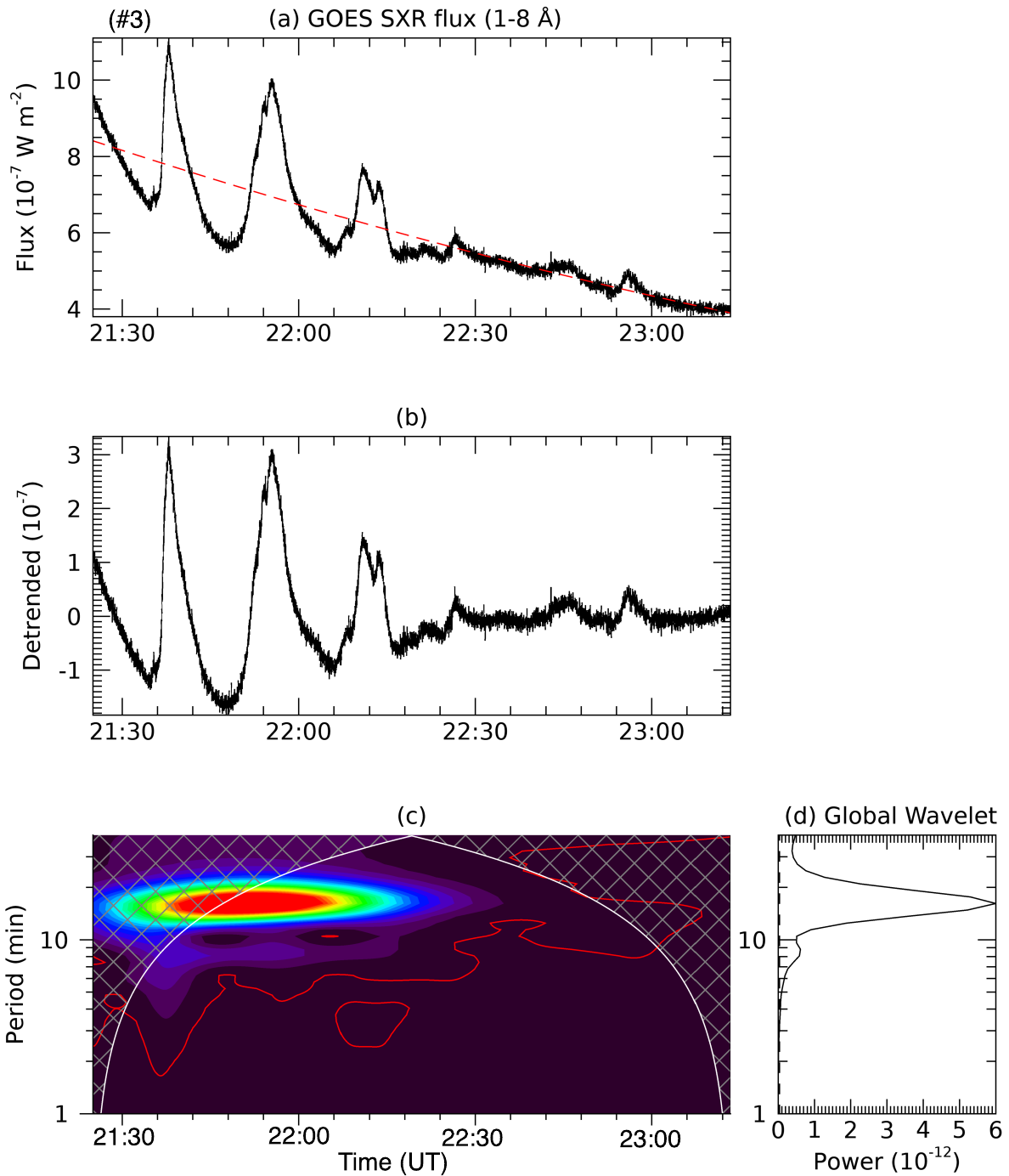}
\includegraphics[width=8.0cm]{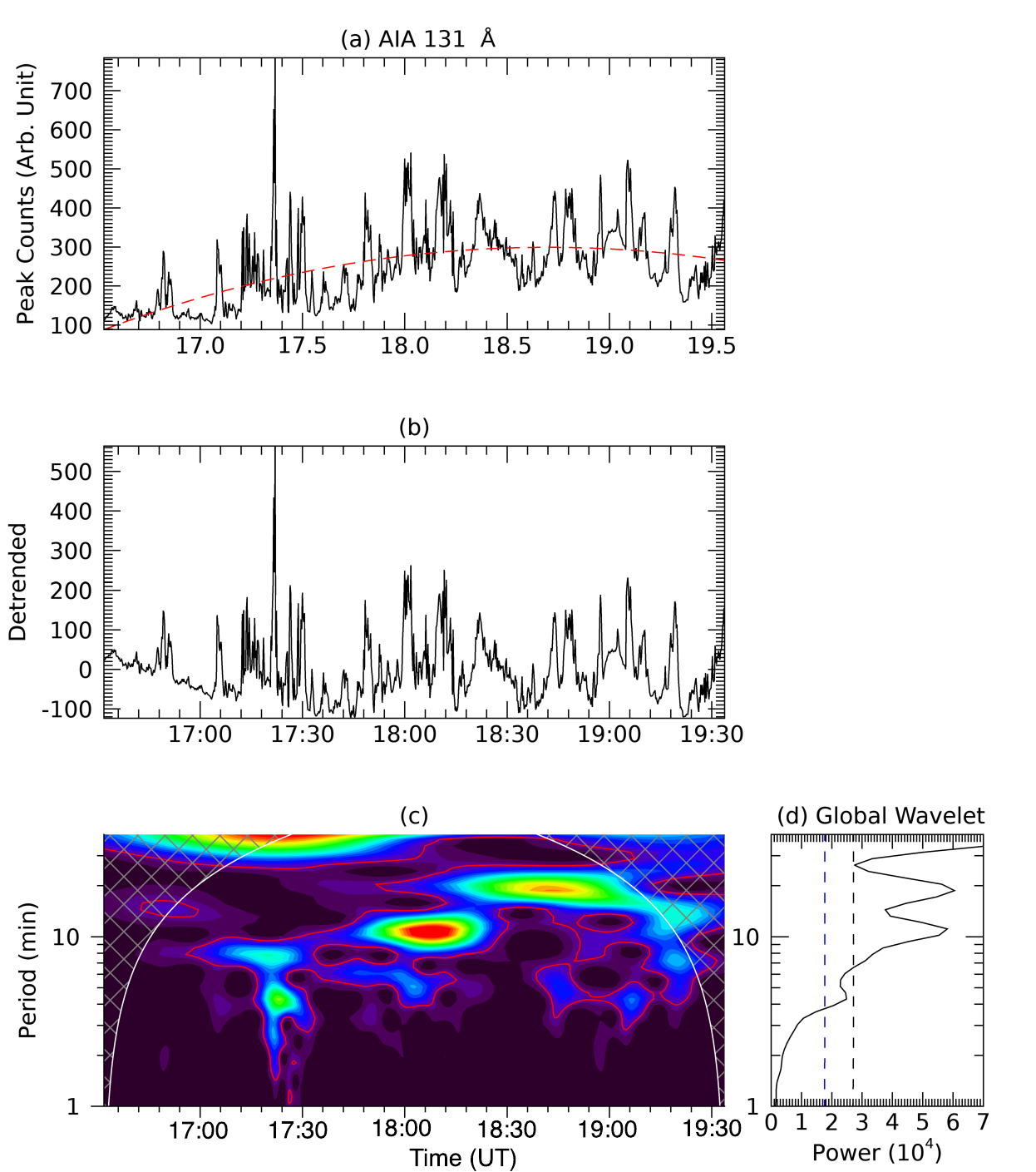}
}
\caption{{\bf Wavelet analyses of the GOES soft X-ray flux and SDO AIA intensity fluctuations.} (a-d) Analysis of X-ray/EUV periodicities in different time intervals. {\it Top left:} (a) Soft X-ray flux in the GOES 1-8 {\AA} channel (\#1). (b) The smoothed and detrended light curve after subtracting the red trend shown in (a) from the original light curve. (c) Wavelet power spectrum of the detrended signal. Red contours outline the 99\% significance level. (d) Global wavelet power spectrum. The dashed line is the 99\% global confidence level. {\it Top right and bottom left:} The same analysis for other intervals \#2 and \#3. {\it Bottom right:} The same analysis for AIA 131 {\AA} intensity variations at the CS. The dashed blue and black lines are the 95\% and 99\% global confidence levels.} 
\label{app-fig4}
\end{figure}
\clearpage
\section {Slow-mode waves}
\begin{figure}[htp]
\centering{
\includegraphics[width=18cm]{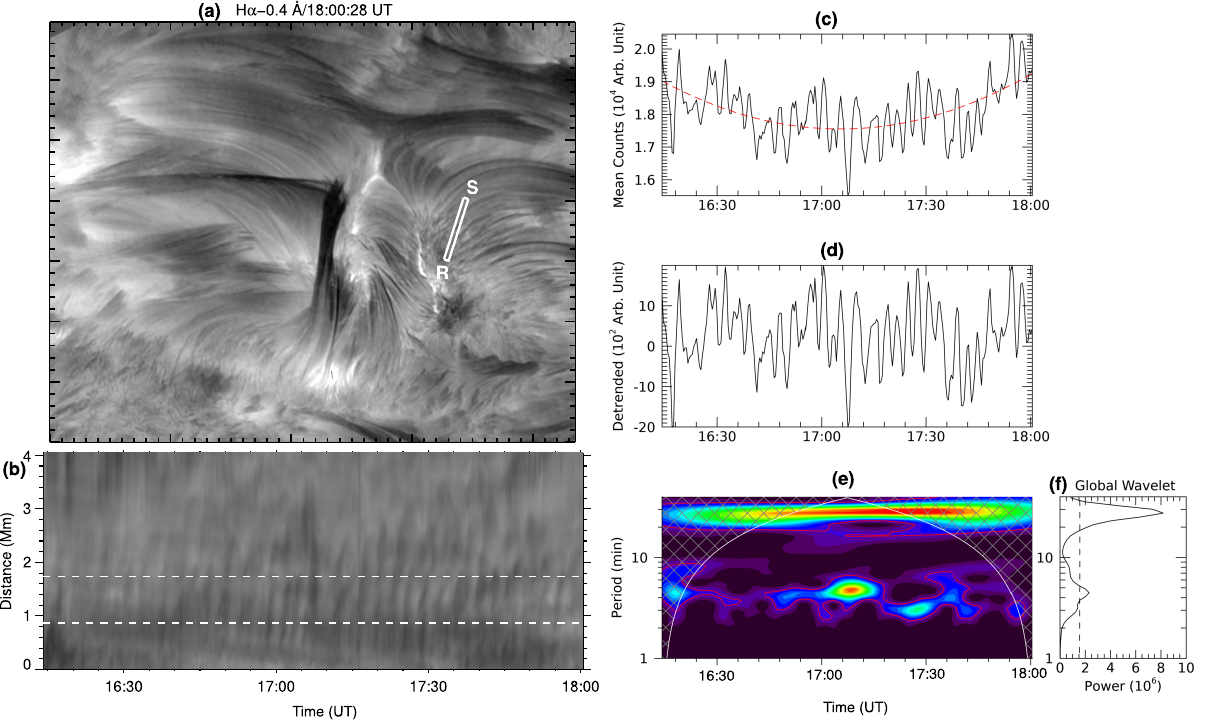}
}
\caption{{\bf Slow magnetoacoustic waves from N2.} (a) H$\alpha$-0.4$~\AA$ image at 18:00:28 UT. Each division on both axes corresponds to a distance of 1$\arcsec$. (b) Time-distance intensity plot along slice RS. (c-f) Wavelet analyses of the intensity of the propagating disturbance (extracted between the two horizontal dashed lines in (b)) from N2. The dashed line is the 95\% global confidence level.} 
\label{app-fig5}
\end{figure}
\clearpage
\section {Breakout model}
\begin{figure}[htp]
\centering{
\includegraphics[width=8cm]{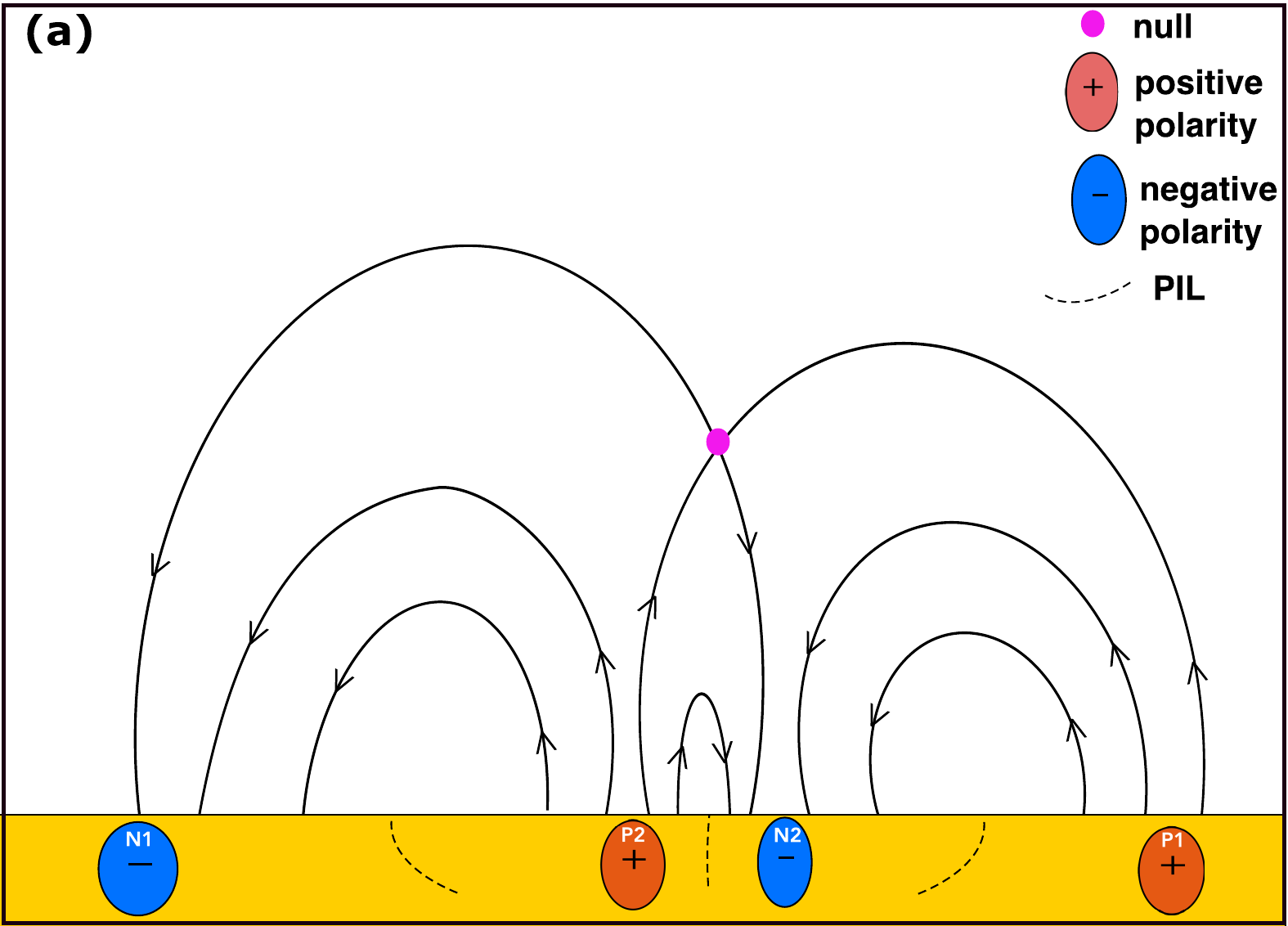}

\includegraphics[width=8cm]{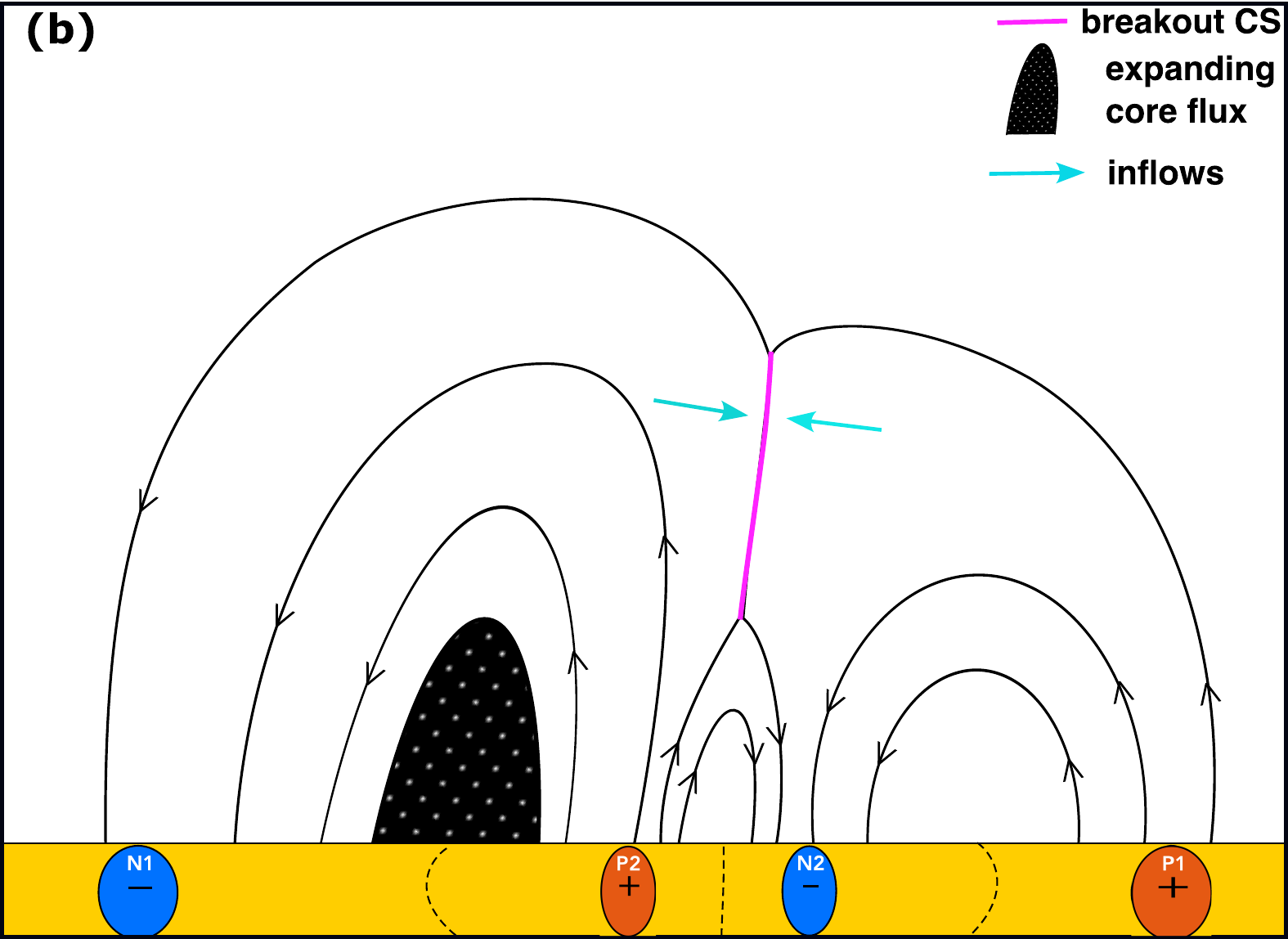}

\includegraphics[width=8cm]{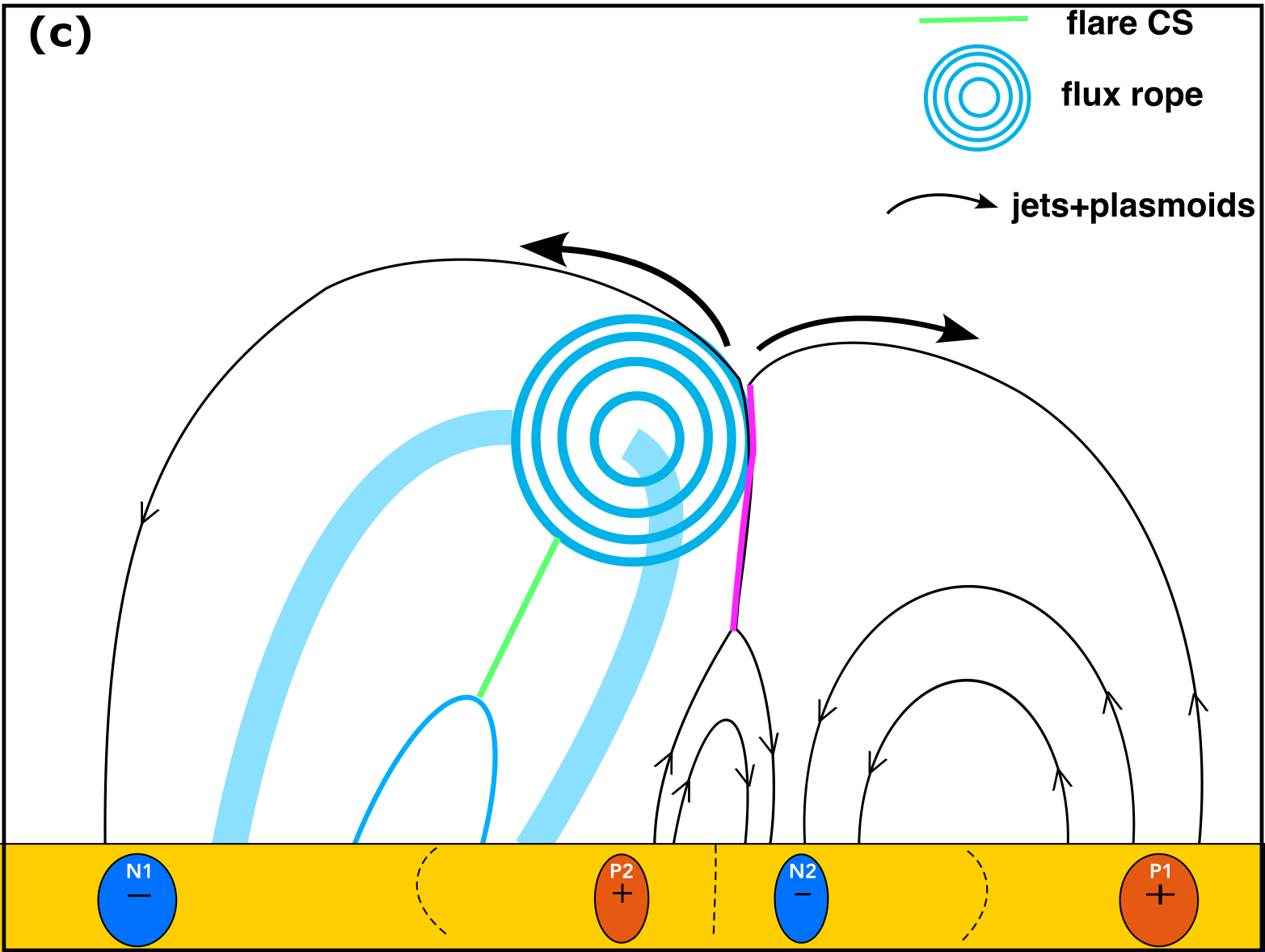}
}
\caption{{\bf Schematic diagram (2D cut) of the three phases of the breakout jet model as applied to the analyzed event.} (a) Pre-eruption fan-spine magnetic configuration. (b) Prolonged modest energy release from breakout reconnection without eruption during B-class flares. Expansion of the core flux (black) above the PIL due to shearing/footpoint motion, and formation of a breakout current sheet (BCS, pink) as the initial null is distorted. Quasiperiodic inflows trigger recurrent reconnection at the BCS and remove the overlying flux above the core flux (filament channel). (c) Intermittent C-class flaring. Flux rope formation and rise due to slow reconnection in the flare current sheet (FCS, green). When the flux rope reaches the BCS, explosive breakout reconnection destroys the flux rope and produces a jet, while fast reconnection in the FCS produces a flare arcade (thin blue loop) and augments the flux rope.} 
\label{app-fig6}
\end{figure}
\clearpage

\section{SUPPLEMENTARY MATERIALS}
\noindent
This section contains supplementary movies to support the results. All high-quality animations are available in the Zenodo repository at : \url{https://doi.org/10.5281/zenodo.12745743}. \\
{\bf Movie S1}: H$\alpha$-0.4 {\AA} animation (Figure \ref{fig3}) and TD intensity plot along P1Q1 (Figure \ref{fig5}(a)) depict the rising structures, inflows, and outflows during prolonged, episodic magnetic reconnection. The animation runs from 16:14 to 18:00 UT. Its real-time duration is 25 s. \\
{\bf Movie S2}: H$\alpha$-0.8 {\AA} animation (Figure \ref{fig4}) and TD intensity plot along P2Q2 (Figure \ref{fig5}(b)) display the rising structures, inflows, and outflows during magnetic reconnection. The animation runs from 16:14 to 18:00 UT. Its real-time duration is 25 s. \\
{\bf Movie S3}: AIA 131 {\AA} animation (Figure \ref{fig4}(g)) and TD intensity plot along P3Q3 (Figure \ref{fig5}(c)) depict inflows, outflows, and associated heating during magnetic reconnection. The animation runs from 16:14 to 18:00 UT. Its real-time duration is 17 s. \\
{\bf Movie S4}: AIA 131 {\AA} animation (Figure \ref{fig4}(h)) and TD intensity plot along P4Q4 (Figure \ref{fig5}(d))  show recurrent brightening at the CS and associated heating during magnetic reconnection. The animation runs from 16:14 to 18:00 UT. Its real-time duration is 15 s. \\
{\bf Movie S5}: Photospheric TiO animation (Figure \ref{app-fig1}(a-e)) displays the evolution of sunspots at the base of the fan-spine topology. The TD intensity plot along the horizontal slit (east to west) depicts the motion of P2$^{\prime}$. The animation runs from 16:14 to 18:00 UT. The second part of the animation shows the evolution of the photospheric magnetic field in HMI magnetograms (Figure \ref{app-fig1}(f-h)) and AIA 304 {\AA} channel images (Figure \ref{fig2}(b)). The animation runs from 15:00 UT to 20:58 UT. Its real-time duration is 20 s. \\
{\bf Movie S6}: AIA 131 {\AA} animation shows the intensity fluctuations at the CS (within the red box, Figures \ref{fig5}(e), \ref{fig6}(a-f, k)), changes in GOES soft X-ray flux (Figure \ref{fig6}(m)), and associated heating during magnetic reconnection. The animation runs from 14:31:47 to 20:57:11 UT. Its real-time duration is 48 s.\\
{\bf Movie S7}: AIA 131 {\AA} animation (Figure \ref{app-fig3} (a-c)) shows long-term evolution of intensity fluctuations at the CS, recurrent flares/jets (GOES soft X-ray flux), and associated plasma heating during magnetic reconnection.  The animation runs from 4:03:18 to 23:48:30 UT. Its real-time duration is 66 s.\\

\clearpage
\end{document}